\documentclass[letterpaper,twocolumn,english,prx,twocolumn,amsmath,amssymb,superscriptaddress]{revtex4-2}
\usepackage[T1]{fontenc}
\UseRawInputEncoding
\usepackage{babel}
\usepackage{prettyref}
\usepackage{amsmath}
\usepackage{amssymb}
\usepackage{graphicx}
\usepackage[bookmarks=true,bookmarksnumbered=false,bookmarksopen=false,
 breaklinks=false,pdfborder={0 0 1},backref=false,colorlinks=false]
 {hyperref}
\hypersetup{pdftitle={Stochastic synaptic dynamics under learning},
 pdfauthor={Jakob Stubenrauch, Naomi Auer, Richard Kempter, and Benjamin Lindner},
 colorlinks=true,linkcolor=black,citecolor=black,urlcolor=black,filecolor=black}

\makeatletter

\usepackage[latin1]{inputenc}

\newrefformat{eq}{\hyperref[#1]{Eq.~(\ref{#1})}}
\newrefformat{cap}{\hyperref[#1]{Fig.~\ref{#1}}}
\newrefformat{fig}{\hyperref[#1]{Fig.~\ref{#1}}}
\newrefformat{tab}{\hyperref[#1]{Table ~\ref{#1}}}
\newrefformat{sec}{\hyperref[#1]{Sec.~\ref{#1}}}
\newrefformat{sub}{\hyperref[#1]{Sec.~\ref{#1}}}
\newrefformat{subsec}{\hyperref[#1]{Sec.~\ref{#1}}}
\newrefformat{cha}{\hyperref[#1]{Chapter~\ref{#1}}}
\newrefformat{appendix}{\hyperref[#1]{Appendix~\ref{#1}}}

\usepackage{MnSymbol}
\DeclareMathAlphabet\mathbb{U}{msb}{m}{n}

\makeatother

\begin{document}
\global\long\def\id{\mathds{1}}%
\global\long\def\epsilon{\varepsilon}%
\global\long\def\cO{\mathcal{O}}%
\global\long\def\cM{\mathcal{M}}%
\global\long\def\cN{\mathcal{N}}%
\global\long\def\da{\text{da}}%
\global\long\def\hyp{_{2}F_{2}}%
\global\long\def\bq{\boldsymbol{q}}%
\global\long\def\bp{\boldsymbol{p}}%
\global\long\def\cD{\mathcal{D}}%

\title{Stochastic synaptic dynamics under learning}
\author{Jakob Stubenrauch}
\affiliation{Bernstein Center for Computational Neuroscience Berlin, 10115 Berlin,
Germany}
\affiliation{Physics Department of Humboldt-Universit\"at zu Berlin, 12489 Berlin,
Germany}
\author{Naomi Auer}
\affiliation{Institute for Theoretical Biology, Department of Biology, Humboldt-Universit\"at
zu Berlin, 10115 Berlin, Germany}
\author{Richard Kempter}
\affiliation{Institute for Theoretical Biology, Department of Biology, Humboldt-Universit\"at
zu Berlin, 10115 Berlin, Germany}
\affiliation{Bernstein Center for Computational Neuroscience Berlin, 10115 Berlin,
Germany}
\affiliation{Einstein Center for Neurosciences Berlin, Berlin 10117, Germany}
\author{Benjamin Lindner}
\affiliation{Bernstein Center for Computational Neuroscience Berlin, 10115 Berlin,
Germany}
\affiliation{Physics Department of Humboldt-Universit\"at zu Berlin, 12489 Berlin,
Germany}
\begin{abstract}
Learning is based on synaptic plasticity, which affects and is driven
by neural activity. Because pre- and postsynaptic spiking activity
is shaped by randomness, the synaptic weights follow a stochastic
process, requiring a probabilistic framework to capture the noisy
synaptic dynamics. We consider a paradigmatic supervised learning
example: a presynaptic neural population impinging in a sequence of
episodes on a recurrent network of integrate-and-fire neurons through
synapses undergoing spike-timing-dependent plasticity (STDP) with
additive potentiation and multiplicative depression. We first analytically
compute the drift- and diffusion coefficients for a single synapse
within a single episode (microscopic dynamics), mapping the true jump
process to a Langevin and the associated Fokker-Planck equations.
Leveraging new analytical tools, we include spike-time--resolving
cross-correlations between pre- and postsynaptic spikes, which corrects
substantial deviations seen in standard theories purely based on firing
rates. We then apply this microdynamical description to the network
setup in which hetero-associations are trained over one-shot episodes
into a feed-forward matrix of STDP synapses connecting to neurons
of the recurrent network (macroscopic dynamics). By mapping statistically
distinct synaptic populations to instances of the single-synapse process
above, we self-consistently determine the joint neural and synaptic
dynamics and, ultimately, the time course of memory degradation and
the memory capacity. We demonstrate that specifically in the relevant
case of sparse coding, our theory can quantitatively capture memory
capacities which are strongly overestimated if spike-time--resolving
cross-correlations are ignored. On the single synapse level, our theory
predicts a transition between acceleration and deceleration of the
synaptic weight, and a possibly non-monotonic dependence of the finite-time
diffusion coefficient on the synaptic weight. On the memory level,
we showcase two complementary roles of homeostatic plasticity, namely
it is both responsible for forgetting and a requirement for the ability
to learn. Additionally, we show how strength of plasticity and duration
of exposure to stimuli must be scaled to evade a tradeoff dilemma
between memory capacity and early accuracy. We conclude with a discussion
of the many directions in which our framework can be extended.
\end{abstract}
\maketitle

\section{Introduction}

Synaptic plasticity constitutes the foundation of learning. Although
the mechanisms responsible for plasticity are diverse and incompletely
understood \citep{Abbott2000_1183,Sjoestroem2010_1362}, the quantitative
dependence of plasticity on the relative timing between pairs of pre-
and postsynaptic spikes \citep{Gerstner1996_78,Markram1997_440,Bi1998_10472,Bi2001_166}
accounts for many of the experimental observations; corrections are
often extensions of this pair-based rule. The statistics of synaptic
weight dynamics is thus determined by the statistics of pre- and postsynaptic
spikes. Because the latter display a considerable variability, synaptic
dynamics has to be described in a stochastic framework, even if the
intrinsic noise sources of plasticity \citep{Yasumatsu2008_13608,Rosenbaum2012_18,Rosenbaum2013_484,Statman2014_17}
are neglected.

Several aspects of synaptic dynamics have been studied before; the
mean dynamics and even ensemble dynamics \citep{Kempter1999_4514,vanRossum2000_8821,Burkitt2004_940,Meffin_2006_041911,Gilson2009_102,Fremaux2016_85,Ocker2018_951,Akil2021_23,Sosis2024_bioRxiv}
have been investigated for a variety of plasticity rules and settings,
see \citep{Sjoestroem2010_1362} for an extensive review. More recently,
the conjunction of classical spike-timing-dependent plasticity (STDP)
with homeostatic plasticity has received increased attention (see
\citep{Keck2017} for an overview).

Classical pair-based STDP implies that the expected synaptic weight
depends on (i) pre- and postsynaptic firing rates and (ii) cross-correlations
between pre- and postsynaptic spikes \citep{Kempter1999_4514}. Since
the weight change induced by STDP is sensitive to relative spike-timing
differences on the order of milliseconds \citep{Markram1997_440,Bi1998_10472,Abbott2000_1183},
these two contributions play fundamentally distinct roles for learning.
The effect of cross-correlations on learning is typically taken into
account by interpreting the post-synaptic neuron as a linear filter.
This somewhat drastic simplification is either explained by model
choice (conditionally Poissonian neuron) or through a daring but educated
trick: to linearize nonlinear neuron models \emph{per realization}
using realization-averaged response functions \citep{Lindner2005_061919,Grytskyy2013_00131,Layer2024_013013},
see e.g. \citep{Gerstner2001610,Ocker2015_40,Ocker2018_951}. Response
functions can be derived with Fokker-Planck theory, for integrate-and-fire
neurons see Refs. \citep{Brunel2001_2189,Lindner2001_2937,Richardson2008,Richardson10_178102,Droste2017_81,Droste2017_012411}.

Despite the vast body of research on synaptic dynamics, a robust and
generalizable stochastic description of synaptic dynamics is still
missing. Such a description is clearly needed to develop multiscale
memory models that take into account the single neuron level and the
population level. In this paper we lay the foundations to this endeavor;
specifically, we analytically characterize the stochastic process
of synapses endowed with STDP that feed Poisson spikes to leaky integrate-and-fire
(LIF) neurons and are shaped by the spikes of both Poisson and LIF
neurons. We focus on the stochasticity due to pre- and postsynaptic
fluctuations and neglect the intrinsic noisiness of synapses. Instead
of relying on realization-wise linearization, we incorporate cross-correlations
through an exact model-independent relation between cross-correlations
and response-functions that two of us reported recently \citep{Stubenrauch2024_041047}.
This approach is robust and generalizable. From the theory developed
here we make several qualitative predictions about synaptic dynamics
and memory.

Synapses endowed with STDP are capable of self-organized formation
of diverse types of memory including (i) learning input correlations
\citep{Guetig2003_3714}, (ii) reinforcement learning in recurrent
networks of LIF neurons \citep{Ocker2018_951} or (iii) error correction
of drifting assemblies \citep{Kossio2021}. Yet another type of STDP-based
memory are hetero-associations, i.e., representations in one neuron
population evoke representations in another one. Building on our results
about stochastic synaptic dynamics, we analytically investigate hetero-associative
memory. As we outline, for characterizing this type of memory, the
stochastic view on synaptic dynamics is indispensable. Extending our
methods to the aforementioned mechanisms (i--iii) poses exciting
problems for future work.

\begin{figure*}
\includegraphics{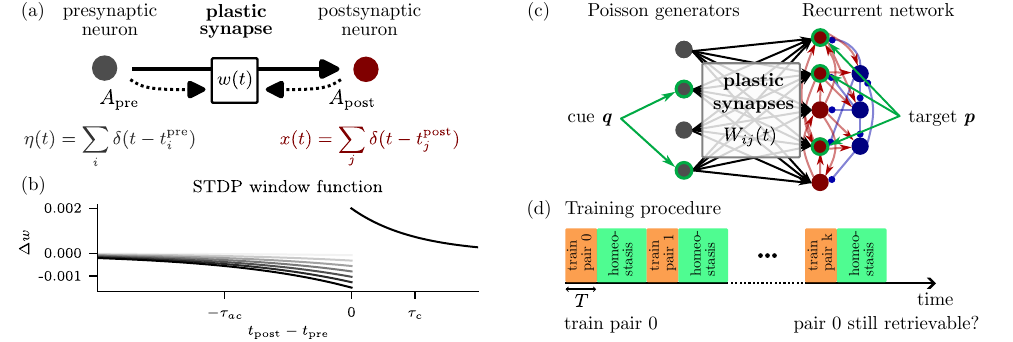}

\caption{\label{fig:joint_model}We investigate memory properties by studying
a single synapse dynamics that effectively captures synaptic population
dynamics. (a) A Poisson process $\eta$ drives the neuron $x$ through
the synapse $w$. The weight $w$ is plastic, i.e., it dynamically
depends on $x$ and $\eta$. We indicated the tracers $A_{\text{pre}}$
and $A_{\text{post}}$ used in the embedding \prettyref{eq:model_exp_local}.
(b) STDP window \prettyref{eq:plasticity_kernel} for $w=0.01$ (light
gray) through $w=0.2$ (black). The setup (a,b) is studied in \prettyref{sec:Stochastic-dynamics-of}
(c) A cue representation $\boldsymbol{q}$ in a layer of Poisson processes
(gray) is activated and drives a recurrent neural network through
a matrix of plastic synapses. Among the excitatory neurons (red) in
the recurrent network, a target representation $\protect\bp$ receives
an additional stimulus. The cue and target subsets are marked by the
green arrows and circles. (d) Training procedure to store hetero-associations
between cues and targets. The setup (c,d) is studied in \prettyref{sec:Dynamics-of-learning}.
Parameters $\Delta_{c}=2\times10^{-3}$, $r_{ac}=8\times10^{-3}$,
$\tau_{c}=16.8/20$, and $\tau_{ac}=33.7/20$ are matched to \citep{Bi1998_10472}.}
\end{figure*}

The paper is organized as follows: In \prettyref{sec:models}, we
first introduce the synapse and neuron model \prettyref{fig:joint_model}(a,b)
studied throughout the paper. Building on these models, we introduce
the network model \prettyref{fig:joint_model}(c) and the training
scheme \prettyref{fig:joint_model}(d), which store hetero-associations
into a plastic feed-forward matrix. In \prettyref{sec:Stochastic-dynamics-of},
we characterize the stochastic process of a single synaptic weight
by deriving its drift- and diffusion coefficients, which define a
Langevin (or corresponding Fokker-Planck) equation. From this description,
we derive the dynamics of the ensemble mean and variance of synaptic
weights. Additionally, we discuss a transition from accelerating to
decelerating synaptic weights, which entails qualitative experimental
predictions. In \prettyref{sec:Dynamics-of-learning}, we leverage
the theory of single weights to study the network scenario in a mean-field
theory of synaptic populations. We apply the mean-field theory to
characterize forgetting and compute the memory capacity of the setup.
Based on this theory, we derive a tradeoff between memory lifetime
and early memory accuracy, which we resolve by an optimal training
scheme. Lastly, we discuss two complementary roles of homeostatic
plasticity.

\section{Models\label{sec:models}}

First, we introduce the synaptic model depicted in \prettyref{fig:joint_model}(a,b).
This model depends on pre- and postsynaptic spike trains, thus, next,
we define the neuron models used in this paper. Then we introduce
the network model depicted in \prettyref{fig:joint_model}(c), which
is composed of the synapse- and neuron models above. We are interested
in how hetero-associations can be stored into this network; the necessary
training scheme {[}see \prettyref{fig:joint_model}(d){]} is explained
last.

\subsection{Synapse model}

We consider a synapse with presynaptic spike train $\eta(t)=\sum_{i}\delta(t-t_{i}^{\text{pre}})$
and postsynaptic spike train $x(t)=\sum_{j}\delta(t-t_{j}^{\text{post}})$.
The evolution of the synaptic weight $w$ in the classical STDP model
\citep{Sjoestroem2010_1362} can be described by 
\begin{align}
\dot{w}(t) & =\int_{-\infty}^{t}dt^{\prime}\,\kappa[t-t^{\prime},w(t)]\eta(t^{\prime})x(t)\nonumber \\
 & \;+\int_{-\infty}^{t}dt^{\prime}\,\kappa[t^{\prime}-t,w(t)]\eta(t)x(t^{\prime}).\label{eq:model_general}
\end{align}
Each pre-post spike pair leads to a jump in $w$ by $\kappa$; causal
spike pairs are captured by the first line, anti-causal pairs by the
second line. Here, $\kappa(\tau,w)$ is the STDP window function,
which we assume only depends on the time difference $\tau$ between
spikes and on $w$ at the time $t$ of update. The lower integral
bounds are set to $-\infty$ so as to avoid switch-on effects of the
plasticity rule. Throughout the paper, we consider an exponential
window with multiplicative (i.e., $w$-dependent) depression and additive
(i.e., $w$-independent) potentiation \citep{vanRossum2000_8821,Morrison2008_478}
\begin{equation}
\kappa(\tau,w)=\begin{cases}
\Delta_{c}e^{-\tau/\tau_{c}} & \tau\geq0\\
-r_{ac}we^{\tau/\tau_{ac}} & \tau<0,
\end{cases}\label{eq:plasticity_kernel}
\end{equation}
see \prettyref{fig:joint_model}(b), with $\Delta_{c},r_{ac}>0$ determining
the amplitudes and $\tau_{c},\tau_{ac}>0$ being the time scales of
potentiation (index $c$ for causal) and depression (index $ac$ for
anticausal), respectively. In addition to Eqs. \eqref{eq:model_general}
and \eqref{eq:plasticity_kernel}, we demand that the considered excitatory
synapse maintains a positive weight throughout, $w(t)\geq0$, a necessary
condition for Dale's law, and implemented in our simulations by a
clipping boundary condition. Since clipping happens very rarely for
the parameters obtained from \citep{Bi1998_10472,Bi2001_166}, we
can safely disregard the clipping rule in the theory. The plasticity
rule \prettyref{eq:plasticity_kernel} with its multiplicative depression
and additive potentiation is chosen as it captures the experimental
findings of Ref. \citep{Bi1998_10472} and because of its inherent
stability. Generalizations of \prettyref{eq:model_general}, e.g.,
triplet rules \citep{Sjoestroem2001_1164,Pfister2006_9682} or neuromodulatory
dynamics \citep{Sosis2024_bioRxiv} are of interest but out of scope
of our study.

For the exponential window in \prettyref{eq:plasticity_kernel}, one
can rewrite \prettyref{eq:model_general} in terms of pre- and postsynaptic
trace variables $A_{\text{pre}}$ and $A_{\text{post}}$ \citep{gerstner1998_377}
\begin{align}
\dot{w} & =\Delta_{c}A_{\text{pre}}x-r_{ac}wA_{\text{post}}\eta,\nonumber \\
\dot{A}_{\text{pre}} & =-\tau_{c}^{-1}A_{\text{pre}}+\eta,\nonumber \\
\dot{A}_{\text{post}} & =-\tau_{ac}^{-1}A_{\text{post}}+x.\label{eq:model_exp_local}
\end{align}
Multiple previous studies introduce their plasticity model in an embedded
form like \prettyref{eq:model_exp_local} (e.g. Refs. \citep{vanRossum2000_8821,Kempter2001_2741,Pfister2006_9682,Clopath2010_352,Zenke2015_6922,Akil2021_23,Sosis2024_bioRxiv})
and interpret the tracer variables $A_{\text{pre}}$ and $A_{\text{post}}$
as concentrations of components relevant for plasticity \citep{gerstner1998_377}.
Besides this potential connection of the phenomenological model \prettyref{eq:model_general}
to the underlying biophysics, this embedding is useful for simulations
as it makes the system local in time; simulation results presented
in this paper are based on integrating \prettyref{eq:model_exp_local},
see \prettyref{appendix:Numerical-methods}. To avoid switch-on effects
in \prettyref{eq:model_exp_local}, we integrate $A_{\text{pre}}$
and $A_{\text{post}}$ for some time before we start integrating $w$;
this way, $A_{\text{pre}}$ and $A_{\text{post}}$ thermalize, i.e.,
their statistics become stationary, which is in line with the lower
integral bound at $-\infty$ in \prettyref{eq:model_general}.

\subsection{Neuron model}

We here introduce a neuron model that is driven (i) by Gaussian white
noise and (ii) by a \emph{single} presynaptic neuron through a single
synapse. The white Gaussian noise is a simple description of unaccounted
noise sources as for instance intrinsic channel noise and synaptic
input from external neural populations that are not explicitly modeled
here. The dynamics of the single synapse will be studied in \prettyref{sec:Stochastic-dynamics-of}.
The dynamics of \emph{multiple} synapses providing input to a neuron,
studied in \prettyref{sec:Dynamics-of-learning}, can be effectively
mapped to the single-synapse case introduced here. The synaptic dynamics
\prettyref{eq:model_general} is driven by the pre- and postsynaptic
neural activity $\eta$ and $x$, respectively. Thus, to specify the
synaptic dynamics, we need to specify the neuron models. Throughout
the paper, $\eta(t)$ is the spike train of a Poisson process with
rate $\nu$, and $x(t)$ is the spike train of a model neuron driven
by $\eta$. The developed framework holds true for arbitrary neuron
models $x$, but we give specific expressions and simulation results
for the LIF neuron $x(t)=\sum_{i}\delta(t-t_{i}^{\text{post}})$,
where $t_{i}^{\text{post}}$ are the times at which the membrane voltage
governed by
\begin{equation}
\tau_{m}\dot{v}=-v+\mu+\sqrt{2D}\xi(t)+w(t)\eta(t)\label{eq:lif}
\end{equation}
hits the threshold $v_{\text{t}}$; additionally, at $t_{i}^{\text{post}}$,
$v$ is reset to $v_{\text{r}}$. Throughout the manuscript, $v_{\text{r}}=0$,
$v_{\text{t}}=1$, and the membrane time constant is set to $\tau_{m}=1$,
i.e., we measure time in multiples of $\tau_{m}$. The input to the
neuron described by \prettyref{eq:lif} consists of the mean current
$\mu$, white Gaussian noise $\sqrt{2D}\xi$ with noise intensity
$D$ and $\left\langle \xi(t)\xi(t^{\prime})\right\rangle =\delta(t-t^{\prime})$,
and the Poissonian spikes weighted by $w$. Later on, when we study
neurons embedded in a recurrent network, the input parameters $\mu$
and $D$ will, in addition to explicit noise, effectively capture
the recurrent input. The LIF neuron is mechanistic enough to generate
spike trains $x$ that are correlated with $\eta$ in a beyond-rate-based
fashion. Moreover, it has been shown to reproduce experimental spike
trains of pyramidal cells \citep{Rauch2003_1612}, and many of its
statistical properties are available \citep{Fourcaud2002_2110,Brunel2001_2189,Lindner2001_2937,Roxin16217}.

\subsection{Network model\label{subsec:Network-model}}

Here, we introduce the network model depicted in \prettyref{fig:joint_model}(c),
which can store hetero-associations, i.e., feed-forward associations
between representations in two distinct populations. A layer of $M$
Poisson processes $\eta_{i}$ (gray disks) drives a recurrent network
of $N$ LIF neurons \citep{Brunel00_183} (red and blue disks) through
a plastic weight matrix $W$ with $W_{ij}$ being the weight by which
the $j$th Poisson process drives the $i$th postsynaptic neuron.
The recurrent network consists of $N_{E}$ excitatory (red) and $N_{I}$
inhibitory (blue) neurons, among which only the excitatory neurons
are targeted by $W$. Additionally, each neuron has exactly $C_{E}$
incoming excitatory connections from the recurrent network and $C_{I}$
incoming inhibitory connections with static weights $J_{EE}=J$, $J_{EI}=J_{II}=-gJ$,
$J_{IE}=hJ$, parameterized by $J>0$, $g>0$, and $h>0$. Specifically,
$g$ ($h$) denotes the ratio of efficacies of I to I (E to I) and
the efficacies of E to E synapses, respectively. In accordance with
neuroanatomical findings, we choose sparse recurrent connectivity
$C_{E},C_{I}\ll N_{E},N_{I}$. The computational advantage of the
recurrent connectivity is to instantiate a competition between neurons
that, in a soft winner-take-all manner, support strongly driven neurons
and suppress weakly driven neurons.

Most input Poisson processes have a low firing rate $\nu_{\text{lo}}$,
however, in each training step, a subset of $f_{c}M$ Poisson generators
(index $c$ for cue neurons), represented by the ones in the binary
vector $\bq\in\left\{ 0,1\right\} ^{M}$ fires with increased rate
$\nu_{\text{hi}}$. Simultaneously, a subset of $f_{s}N_{E}$ excitatory
neurons (index $s$ for supervision), represented by the binary vector
$\bp\in\left\{ 0,1\right\} ^{N_{E}}$, receives direct Poissonian
input $\gamma_{i}$ with rate $\nu_{s}$ and static weight $J_{s}$;
the other $(1-f_{s})N_{E}$ excitatory neurons do not receive additional
input. The neurons stimulated by cue and supervision are marked by
green circles in \prettyref{fig:joint_model}. We are particularly
interested in sparse representations, where the activation ratios
$f_{c},f_{s}\ll1$.

Summarizing the above, the membrane voltages of the excitatory neurons
follow 
\begin{align}
\dot{v}_{i} & =-v_{i}+\mu_{E}+\sqrt{2D_{E}}\xi_{i}(t)+J\sum_{n\in C_{E}(i)}x_{n}-gJ\sum_{n\in C_{I}(i)}x_{n}\nonumber \\
 & \quad+\sum_{j}W_{ij}(t)\eta_{j}(t)+p_{i}J_{s}\gamma_{i}(t),\quad\quad i=1,...,N_{E}\label{eq:brunel_net_E-1}
\end{align}
and the membrane voltages of the inhibitory neurons obey 
\begin{align}
\dot{v}_{i} & =-v_{i}+\mu_{I}+\sqrt{2D_{I}}\xi_{i}(t)+hJ\sum_{n\in C_{E}(i)}x_{n}-gJ\sum_{n\in C_{I}(i)}x_{n},\nonumber \\
 & \quad i=N_{E}+1,...,N_{E}+N_{I}\label{eq:brunel_net_I-1}
\end{align}
where the spike trains $x_{n}(t)=\sum_{i}\delta(t-t_{n,i})$ are given
in terms of the fire-and-reset times $t_{n,i}$ of neuron $n$ {[}see
\prettyref{eq:lif}{]}, and $\mu_{E},D_{E}$ ($\mu_{I},D_{I}$) are
the baseline mean input and noise-intensity of the excitatory (inhibitory)
neurons. The white noise contributions aim to capture noise sources
such as channel noise; when approximating Eqs.~\eqref{eq:brunel_net_E-1}
and \eqref{eq:brunel_net_I-1} with effective instances of the single
neuron \prettyref{eq:lif}, the total noise intensity $D$ consists
of the explicit noise $D_{E/I}$ and the noise intensity of the network
and Poissonian input. The recurrent network is a variant of \citep[Model A with $h=1$]{Brunel00_183}.
For $h\neq1$, excitatory spikes have different efficacies at excitatory
and inhibitory neurons, respectively. For $h>1$, this establishes
a competition between the neurons that proves useful for long memory.
In a study of rat neocortex, $J_{IE}\approx2J_{EE}$ (i.e., $h=2$)
has been reported for regular spiking excitatory neurons and fast
spiking inhibitory interneurons \citep{Beierlein2003_3000,Bernardi2021_38};
throughout the paper we use $h=2$ exclusively.

\subsection{Training scheme\label{subsec:Training-procedure-1}}

The training procedure described here mimics the situtation where
a pre-synaptic assembly (a \emph{cue}) is activated and a supervisor
drives a \emph{target} assembly in the post-synaptic population. Then,
through STDP, the matrix $W$ learns to associate the cue with the
target: after some time, activating the cue will autonomously activate
the target without requiring the supervisor---the hetero-association
is stored.

Specifically, the goal of training is to store associations of random
(independently drawn) pattern pairs $(\bq_{k},\bp_{k})$ into the
network \prettyref{fig:joint_model}(c). The training procedure is
illustrated in \prettyref{fig:joint_model}(d) and proceeds as follows.
The zeroth association is trained by setting $\bq=\bq_{0}$ and $\bp=\bp_{0}$.
First, trace variables and membrane voltages are integrated for $T_{\text{warm}}=20$
with frozen $W$. Second, all dynamical variables, including the entries
of $W$, are integrated for a time $T$; here, the synapses $W_{ij}$
follow the dynamics \prettyref{eq:model_general} with the specific
kernel \prettyref{eq:plasticity_kernel}, see \prettyref{appendix:Numerical-methods}
for implementation details. We assume that there is a pause before
the next pattern pair is trained. During this pause, homeostatic plasticity
occurs, which, according to experimental findings, can be modeled
as a slow rescaling of weights to maintain firing rates (see \citep{Zenke2017_20160259,Keck2017}
and references therein) or summed synaptic weight \emph{per postsynaptic
neuron }(approximating the experimentally observed conservation of
summed synapse surface area per postsynaptic neuron over time, see
\citep{Bourne2011_373}). We follow the latter view, and thus, after
each training session, we rescale each weight as 
\begin{equation}
W_{ij}\rightarrow W_{ij}\frac{m_{0}}{M^{-1}\sum_{j^{\prime}=1}^{M}W_{ij^{\prime}}},\label{eq:homeostasis-1}
\end{equation}
where we introduced the parameter $m_{0}$ that defines the average
synaptic weight per postsynaptic neuron. Note that each row of $W$
sums to $Mm_{0}$. After homeostasis, we set $\bq=\bq_{1}$ and $\bp=\bp_{1}$,
and repeat the entire procedure, then proceed with $(\bq_{2},\bp_{2})$
and so on.

\begin{figure*}
\includegraphics{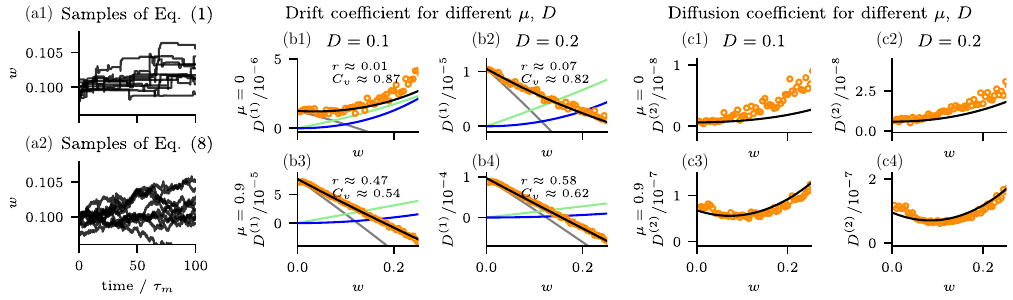}

\caption{\label{fig:single_synapse_full}Stochastic dynamics of single synapses.
(a1) Sample trajectories of \prettyref{eq:model_general} and (a2)
sample trajectories of the corresponding Langevin equation \prettyref{eq:langevin}
for $\mu=0.6$, $D=0.2$, $\nu=0.1$, $m_{0}=0.1$, $\sqrt{V_{0}}=10^{-3}$
and STDP parameters as in \prettyref{fig:joint_model}(b). (b1--b4)
Drift coefficient $D^{(1)}$ from simulations (orange circles) and
theory {[}\prettyref{eq:D1} (black line) and single contributions:
firing-rate (gray), mean-response (green), and noise-intensity--response
(blue){]} for different $\mu$ and $D$. The instantaneous firing
rate $r$ and the coefficient of variation $C_{v}$ at $w=0.1$ are
indicated in the upper center of the four panels. (c1--c4) Finite-time
diffusion coefficient $D^{(2)}$ for $\Delta t=10$. Simulation results
(orange circles). Theory (black lines) $[V(\Delta t)+(m(\Delta t)-w)^{2}]/(2\Delta t)$,
with $m$ from \prettyref{eq:solution_mean_drift} and $V$ from \prettyref{eq:solution_diffusion}.}
\end{figure*}

\section{Stochastic synaptic dynamics\label{sec:Stochastic-dynamics-of}}

The dynamics of a synaptic weight $w$, \prettyref{eq:model_general},
is stochastic due to the randomness of $\eta$ and $x$. Trajectories
of $w$ can be obtained in simulations, for instance, for the window
function \prettyref{eq:plasticity_kernel}, one has to integrate both
stochastic equations, Eqs. \eqref{eq:model_exp_local} and \eqref{eq:lif}.
The synaptic weight performs a jump process \prettyref{fig:single_synapse_full}(a1):
with each new pre- or post-synaptic spike, a new set of spike pairs
is formed that leads to a finite jump, and between jumps the weight
is constant. In this section, we develop a simplified description
in terms of a Langevin equation and use it to calculate the transient
mean and the variance of an ensemble of synapses.

The rate of change of synaptic weight is given by \prettyref{eq:model_general}.
Assuming that the amplitudes $\kappa$ are sufficiently small and
the rate of spike pairs sufficiently high, $\dot{w}$ can be approximated
by a white Gaussian process. This leads to a Langevin equation
\begin{equation}
\dot{w}=D^{(1)}(w)+\sqrt{2D^{(2)}(w)}\zeta(t),\label{eq:langevin}
\end{equation}
which, due to causality of the synaptic updates, we interpret in the
sense of Ito \citep{gardiner1985handbook}. Here, $\zeta(t)$ is white
Gaussian noise with $\left\langle \zeta(t)\right\rangle =0$ and $\left\langle \zeta(t)\zeta(t^{\prime})\right\rangle =\delta(t-t^{\prime})$.
Samples of \prettyref{eq:langevin} are shown in \prettyref{fig:single_synapse_full}(a2).
Correspondingly, the transition probability $p(w,t|w_{0},t_{0})$
of weights follows the Fokker-Planck equation 
\begin{align}
\partial_{t}p(w,t|w_{0},t_{0}) & =-\partial_{w}D^{(1)}(w)p(w,t|w_{0},t_{0})\nonumber \\
 & \quad+\partial_{w}^{2}D^{(2)}(w)p(w,t|w_{0},t_{0}).\label{eq:fokker_planck}
\end{align}
In this section, we derive the functions $D^{(1)}(w)$ and $D^{(2)}(w)$,
which we then apply to describe the evolution of an ensemble of synaptic
weights.

\subsection{Drift coefficient}

The drift coefficient $D^{(1)}$ is the first Kramers-Moyal coefficient
\begin{equation}
D^{(1)}(w)=\lim_{\Delta t\rightarrow0}\frac{1}{\Delta t}\left\langle w_{\text{traj}}(t+\Delta t)-w_{\text{traj}}(t)\right\rangle _{w_{\text{traj}}(t)=w},\label{eq:drift_def.}
\end{equation}
where $w_{\text{traj}}$ is a sample of the stochastic process \prettyref{eq:model_general},
and the subscript denotes a condition on the ensemble average. Assuming
slow weight dynamics (compared to the neuron's relaxation time scale
and to the width of the STDP kernel, detailed in \prettyref{appendix:drift_by_cross_corr}),
\prettyref{eq:drift_def.} can be expressed by stationary statistics
of the LIF neuron
\begin{equation}
D^{(1)}(w)=\int_{-\infty}^{\infty}d\tau\,\kappa(\tau,w)\left[\nu r(w)+C_{x\eta}(\tau,w)\right],\label{eq:force}
\end{equation}
which is a well known result \citep{Kempter1999_4514}. Here, $r(w)\equiv\left\langle x\right\rangle $
is the instantaneous firing rate of the postsynaptic neuron assuming
weight $w$, and $C_{x\eta}(\tau,w)=\left\langle x(t+\tau)\eta(t)\right\rangle _{w}-r(w)\nu$
is the input-spikes--output-spikes cross-correlation. Within a diffusion
approximation of the input to the LIF neuron
\begin{align}
\mu+\sqrt{2D}\xi(t)+w\eta(t)\approx\mu_{\text{DA}}+\sqrt{2D_{\text{DA}}}\xi(t),\label{eq:diff_approx_input}
\end{align}
where $\mu_{\text{DA}}=\mu+w\nu$ and $D_{\text{DA}}=D+w^{2}\nu/2$,
the instantaneous firing rate is given by \citep{Fourcaud2002_2110}
\begin{equation}
\frac{1}{r(w)}=\int_{\frac{v_{r}-\mu_{\text{DA}}}{\sqrt{2D_{\text{DA}}}}}^{\frac{v_{t}-\mu_{\text{DA}}}{\sqrt{2D_{\text{DA}}}}}ds\,e^{s^{2}}[1+\text{erf}(s)],\label{eq:siegert}
\end{equation}
see \citep{Layer2022_5196} for an efficient and stable evaluation
of \prettyref{eq:siegert}. For the STDP kernel in \prettyref{eq:plasticity_kernel},
the first, firing-rate--based, part in \prettyref{eq:force} is

\begin{equation}
\int_{-\infty}^{\infty}d\tau\,\kappa(\tau,w)\nu r(w)=(\Delta_{c}\tau_{c}-r_{ac}w\tau_{ac})\nu r(w).
\end{equation}

The interesting part in \prettyref{eq:force} from a spike-coding
perspective is the cross-correlation $C_{x\eta}$. Since $\eta$ is
assumed to be Poissonian, the cross-correlation is (despite the nonlinearity
of the neuron model) exactly related to the firing-rate response \citep{Stubenrauch2024_041047}
\begin{equation}
C_{x\eta}(\tau)=\nu\frac{\delta}{\delta\nu(t)}\left\langle x(t+\tau)\right\rangle ,\label{eq:crr}
\end{equation}
where $\frac{\delta}{\delta\nu(t)}f[\nu]\equiv\lim_{h\rightarrow0}\frac{d}{dh}f[\nu+h\delta(t-\circ)]$
denotes a functional derivative (here, $\circ$ represents the time
argument of $\nu$ in $f$). The response function of the output spikes
of a LIF neuron to \emph{rate} modulations is to our knowledge not
known (but see \citep{Richardson10_178102} for such a result if the
input amplitudes are exponentially distributed). However, in the diffusion
approximation of the LIF neuron's input in \prettyref{eq:diff_approx_input},
one can apply the chain rule to obtain 
\begin{equation}
\frac{\delta\left\langle x(t+\tau)\right\rangle }{\delta\nu(t)}\approx w\frac{\delta\left\langle x(t+\tau)\right\rangle }{\delta\mu_{\text{DA}}(t)}+\frac{1}{2}w^{2}\frac{\delta\left\langle x(t+\tau)\right\rangle }{\delta D_{\text{DA}}(t)}.
\end{equation}
Thus, the response function in \prettyref{eq:crr} can be approximated
in terms of the response functions to mean- and to noise-intensity
modulations. Their Fourier transforms, $\alpha(\Omega)$ (the susceptibility
to mean modulations) and $\beta(\Omega)$ (the susceptibility to noise-intensity
modulations), can be derived with Fokker-Planck theory and are known
in terms of special functions \citep{Brunel2001_2189,Lindner2001_2937},
also presented in \prettyref{appendix:Response_LIF}. For the exponential
STDP kernel in \prettyref{eq:plasticity_kernel}, the integral in
\prettyref{eq:force} depends on the Laplace transform of the response
functions evaluated at $1/\tau_{c}$ {[}note that due to causality,
$C_{x\eta}(\tau<0)=0${]}; since the Fourier transforms $\alpha$
and $\beta$ are analytic, we may simply evaluate them at $i/\tau_{c}$.
Thus, summing up, the drift coefficient is
\begin{align}
D^{(1)}(w) & =(\Delta_{c}\tau_{c}-r_{ac}w\tau_{ac})\nu r\nonumber \\
 & \;+\Delta_{c}\nu\left[w\alpha(i\tau_{c}^{-1})+\frac{1}{2}w^{2}\beta(i\tau_{c}^{-1})\right].\label{eq:D1}
\end{align}
$D^{(1)}$ is shown and dissected into different contributions in
\prettyref{fig:single_synapse_full}(b) and agrees with simulations
of the system. The drift as a function of $w$ may exhibit a sign
change (b3,b4), in line with experimental observations \citep[Fig.\,2]{Statman2014_17}.
For small weights, the drift is dominated by the firing-rate contributions
{[}first line in \prettyref{eq:D1} and gray lines in \prettyref{fig:single_synapse_full}(b){]},
whereas for increasingly larger weights the response contributions
dominate the drift. The mean-response contribution $\Delta_{c}\nu w\alpha(i\tau_{c}^{-1})$
and the noise-intensity--response contribution $\Delta_{c}\nu(1/2)w^{2}\beta(i\tau_{c}^{-1})$
are shown in \prettyref{fig:single_synapse_full}(b) by the green
and the blue line, respectively. When computing the drift using realization-wise
linearization $x\approx x_{0}+\alpha\ast\eta$ for calculating the
cross correlation function \citep{Lindner2005_061919}, the noise-intensity
response is missing. \prettyref{eq:D1} seems to be accurate for multiple
regimes, including low firing rates and irregular spiking {[}note
the wide range of firing rates and the large coefficients of variation
$C_{v}$ in the examples depicted in \prettyref{fig:single_synapse_full}(b){]}.
However, for low firing rates and large weights $w$ leading to large
synaptic depression jumps, the synaptic jump process is far from a
Langevin process; here \prettyref{eq:D1} deviates from simulation
results of the true jump process. Additionally, for very small intensity
of the white Gaussian noise $D\lesssim10^{-2}$ in \prettyref{eq:lif},
the diffusion approximations required for the firing rate and the
response functions break down which renders \prettyref{eq:D1} inaccurate,
see \prettyref{appendix:Breakdown} for details. For other neuron
models than the LIF neuron, one only needs to substitute the respective
mean- and noise-intensity responses.

\subsection{Diffusion coefficient}

The diffusion coefficient is the second Kramers-Moyal coefficient
\begin{align}
D^{(2)}(w) & =\frac{1}{2}\lim_{\Delta t\rightarrow0}\frac{1}{\Delta t}\nonumber \\
 & \times\left\langle \left[w_{\text{traj}}(t+\Delta t)-w_{\text{traj}}(t)\right]^{2}\right\rangle _{w_{\text{traj}}(t)=w},\label{eq:D2_def}
\end{align}
where the average is over sample trajectories $w_{\text{traj}}$ of
\prettyref{eq:model_general}. Since the weight involves integrals
over the product of $x$ and $\eta$, the squared weights in \prettyref{eq:D2_def}
lead to four-point correlation functions. This makes it difficult
to exactly evaluate \prettyref{eq:D2_def} in general. However, the
diffusion on time scales $\Delta t$ above the inverse firing rates
may be estimated by applying Wick's theorem, as an approximation since
$(\eta,x)$ is a non-Gaussian process, to the four-point correlation
functions, which leads to (see \prettyref{appendix:Noise-intensity-wdot}
for details)
\begin{equation}
D^{(2)}(w)\approx\frac{1}{4}r\nu(\Delta_{c}^{2}\tau_{c}+r_{ac}^{2}w^{2}\tau_{ac}).\label{eq:D2}
\end{equation}
\prettyref{eq:D2} is the infinitesimal diffusion coefficient. For
non-infinitesimal $\Delta t$ long enough (specifically, in \prettyref{fig:single_synapse_full}(c),
$\Delta t=10$), the diffusion is captured by the Langevin dynamics
based on Eqs. \eqref{eq:D1} and \eqref{eq:D2}, as we derive in the
next section.

\subsection{Mid- and long-term evolution of an ensemble of synapses\label{subsec:solving_single_synapse}}

The drift- and diffusion coefficients derived above describe the process
on infinitesimal times. Here, we study the mean and variance of an
ensemble of synapses on finite times. We provide a technically detailed
analysis because the results of this section are the main building
blocks of the mean-field theory of learning discussed in the next
section (\prettyref{sec:Dynamics-of-learning}). Although the dynamics
can admit a stationary solution {[}see e.g. the zero of $D^{(1)}(w)$,
i.e., the black solid line crosses zero in \prettyref{fig:single_synapse_full}(b3-b4){]},
we are here mainly interested in transient dynamics: If the training
patterns permanently (but slowly) change and the synapse is subject
to slow homeostatic plasticity, the stationary state will never be
reached in a learning situation.  Therefore we here study the non-equilibrium
dynamics of the ensemble mean and variance.

We start the discussion from a general point of view and recover that
the drift- and diffusion coefficients derived above suffice for our
purpose. The transition probability $p$ of a Markov process obeys
the Kramers-Moyal expansion
\begin{equation}
\frac{\partial}{\partial t}p(w,t|w^{\prime},t^{\prime})=\sum_{k=1}^{\infty}\left(-\frac{\partial}{\partial w}\right)^{k}\left[D^{(k)}(w)p(w,t|w^{\prime},t^{\prime})\right],\label{eq:kramers_moyal_expansion}
\end{equation}
with the Kramers-Moyal coefficients 
\begin{align}
D^{(k)}(w) & =\frac{1}{k!}\lim_{\Delta t\rightarrow0}\frac{1}{\Delta t}\nonumber \\
 & \times\left\langle \left[w_{\text{traj}}(t+\Delta t)-w_{\text{traj}}(t)\right]^{k}\right\rangle _{w_{\text{traj}}(t)=w}.\label{eq:kramers_moyal_coeffs}
\end{align}
From \prettyref{eq:kramers_moyal_expansion}, we can derive the evolution
of moments $\left\langle w^{n}\right\rangle $ in terms of Kramers-Moyal
coefficients. To this end, we multiply \prettyref{eq:kramers_moyal_expansion}
with $w^{n}$ and integrate over $w$. On the left-hand-side we have
the $n$th moment's time derivative $\partial_{t}\left\langle w^{n}\right\rangle $
under the condition $w^{\prime},t^{\prime}$. On the right hand side,
each summand can be understood using integration by parts: the expression
\begin{align*}
\int w^{n}\left(-\frac{\partial}{\partial w}\right)^{k}\left[D^{(k)}(w)p(w,t|w^{\prime},t^{\prime})\right]dw
\end{align*}
equals a sum of boundary terms plus (if $n\ge k$) $\left\langle n!/(n-k)!w^{n-k}D^{(k)}\right\rangle $.
It is in our case reasonable to assume that the products $D^{(k)}(w)p(w,t|w^{\prime},t^{\prime})$
are vanishing and flat at the boundaries $w\in\{0,\infty\}$ and thus
boundary terms do not contribute, leaving us with \citep{Rahimi_Tabar2019_book}
\begin{equation}
\partial_{t}\left\langle w^{n}\right\rangle =\sum_{k=1}^{n}\frac{n!}{(n-k)!}\left\langle w^{n-k}D^{(k)}(w)\right\rangle .\label{eq:moments_from_km-1}
\end{equation}
The Langevin approximation, which only considers the first two Kramers-Moyal
coefficients, thus implies
\begin{align}
\dot{m} & =\left\langle D^{(1)}(w)\right\rangle \nonumber \\
\dot{V} & =2\left[\left\langle wD^{(1)}(w)\right\rangle -\left\langle w\right\rangle \left\langle D^{(1)}(w)\right\rangle \right]+2\left\langle D^{(2)}(w)\right\rangle ,\label{eq:mV_evolve_exact}
\end{align}
where $m=\left\langle w\right\rangle $ and $V=\left\langle (w-\left\langle w\right\rangle )^{2}\right\rangle $
are the ensemble mean and variance. One may Taylor expand $D^{(1)}$
and $D^{(2)}$ around the self-consistent solution $m(t)$; if $D^{(1)}$
is sufficiently smooth, this leads to
\begin{align}
\dot{m} & =D^{(1)}(m)\label{eq:solution_mean_drift}\\
\dot{V} & =2D^{(1)\prime}(m)V+\frac{r(m)\nu}{2}\left[\Delta_{c}^{2}\tau_{c}+r_{ac}^{2}\tau_{ac}\left(V+m^{2}\right)\right].\label{eq:ode_variance}
\end{align}
We evaluate the mean dynamics \prettyref{eq:solution_mean_drift}
with a Runge-Kutta scheme; \prettyref{eq:ode_variance} can then be
integrated and yields
\begin{align}
V & =e^{A(t)}\left(V_{0}+\int_{0}^{t}dt^{\prime}2D^{(2)}[m(t^{\prime})]e^{-A(t^{\prime})}\right)\nonumber \\
A(t) & =\int_{0}^{t}2D^{(1)\prime}[m(t^{\prime})]dt^{\prime}+\frac{t}{2}r\nu r_{ac}^{2}\tau_{ac}.\label{eq:solution_diffusion}
\end{align}
Equations \eqref{eq:solution_mean_drift} and \eqref{eq:solution_diffusion}
can be efficiently evaluated to obtain the mean and variance of an
ensemble of synapses. These equations are thus useful formulations
of the microscopic (i.e., single-synapse) model, which we leverage
in the next section to derive macroscopic (i.e., population) dynamics.
We also use Eqs. \eqref{eq:solution_mean_drift} and \eqref{eq:solution_diffusion}
to verify the second moment of the stochastic increment over a finite
time window \prettyref{fig:single_synapse_full}(c); i.e., due to
the invalidity of the infinitesimal diffusion coefficient, we predict
the finite-time diffusion coefficient as put forward in \citep{Ragwitz2001_254501}.
For low mean input $\mu$ and thus low firing rates, the finite-time
diffusion coefficient is monotonically growing in $w$, see \prettyref{fig:single_synapse_full}(c1,c2),
in line with the infinitesimal diffusion coefficient \prettyref{eq:D2}.
However, for larger $\mu$ and thus larger rates, \prettyref{fig:single_synapse_full}(c3,c4),
the finite-time diffusion coefficient is non-monotonic and minimal
at a non-vanishing weight $w$. Experimentally, this could be tested
by increasing an optogenetic stimulus (roughly, increasing $\mu$)
and observing if indeed the diffusion becomes non-monotonic.

We lastly add a remark on the truncation of the stochastic process
to only two Kramers-Moyal coefficients (Fokker-Planck level). To this
end, we reflect on \prettyref{eq:moments_from_km-1}. While explicitly
the evolution of the $n$th moment only depends on the first $n$
Kramers-Moyal coefficients, it implicitly depends on the full expansion
through the probability density defining the expectation value. However,
when the Kramers-Moyal coefficients are smooth enough and the spread
of the ensemble small enough such that $D^{(n)}$ is well captured
by its $n$th order Taylor approximation, then the evolution of the
$n$th moment only depends on moments up to order $n$. Consequently,
in this case we achieve moment closure at order $n$, i.e. the system
is self-consistently described by the dynamics of the first $n$ moments.

\subsection{Acceleration or deceleration of synaptic weights?\label{subsec:Accelerate?}}

\begin{figure}
\includegraphics{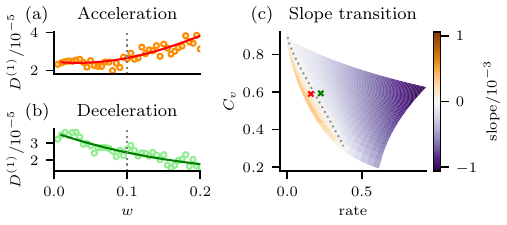}

\caption{\label{fig:acceleration_transition}Acceleration-to-deceleration transition.
(a) Drift coefficient, theory (red line) and simulation (orange circles),
for $\mu=0.75$, $D=0.028$, $r\approx0.16$, $C_{v}\approx0.58$,
corresponding to the red cross in (c). (b) Drift coefficient, theory
(green line) and simulation (light-green circles), for $\mu=0.71$,
$D=0.057$, $r\approx0.23$, $C_{v}\approx0.59$, corresponding to
the green cross in (c). (c) Slope of $D^{(1)}(w=0.1)$ obtained from
\prettyref{eq:D1} plotted against the rate and $C_{v}$ of the postsynaptic
neuron assuming $w=0.1$ fixed. Other parameters as in \prettyref{fig:single_synapse_full}.}
\end{figure}

In \prettyref{fig:single_synapse_full}(b) we observed acceleration
{[}$D^{(1)}(w)$ grows with $w${]} and deceleration {[}$D^{(1)}(w)$
decreases with $w${]} of the synaptic weight. Both, acceleration
and deceleration of synaptic weights seem reasonable. Acceleration
seems reasonable, because the potentiation of a synaptic weight increases
the chance of causal spike pairs. Deceleration seems reasonable as
depression scales more strongly with synaptic weight than potentiation
\citep{Bi1998_10472}, here modeled by the multiplicative nature of
depression, while potentiation is additive {[}\prettyref{eq:plasticity_kernel}{]}.
Eventually, at large $w$, all weights must suffer a deceleration.
To see this, consider the case $w\gg1$. At this point, increasing
$w$ cannot strengthen the postsynaptic neuron's response anymore--already,
one presynaptic spike is guaranteed to cause a postsynaptic spike.
Thus, a further increase of $w$ does not increase the chance of causal
(i.e., potentiating) spike pairs. Yet, both causal (potentiating)
and anticausal (depressing) spike pairs may occur by chance, the latter
though being much more drastic in effect as depression scales with
$w$. Returning to more reasonable weights $w\lesssim0.2$ (i.e.,
more than five spikes needed to carry from reset voltage to threshold
voltage), the drift coefficient can indeed change the sign of its
slope. In \prettyref{fig:acceleration_transition}(a,b), we show the
drift coefficient for the two cases of acceleration and deceleration.

How does the transition from acceleration to deceleration depend on
the dynamics of the postsynaptic cell? In order to investigate this
question, we vary the mean input $\mu$ and the noise intensity $D$
of the LIF neuron. Assuming a fixed weight (here, $w=0.1$), $\mu$
and $D$ uniquely determine the firing rate and the coefficient of
variation $C_{v}$ of the postsynaptic neuron \citep{Vilela2009_90}.
In \prettyref{fig:acceleration_transition}(c), we show the slope
of $D^{(1)}(w=0.1)$ color-coded in the plane of postsynaptic rate
and $C_{v}$. We observe that acceleration occurs only in the lower
left corner of the rate-$C_{v}$ plane (i.e., for small rates and
a $C_{v}$ almost linearly bound by the rate). The picture does not
change qualitatively when varying the reference point $w$ or input
rate $\nu$.

Fluorescence experiments estimating spine-volume dynamics {[}a proxy
for $w(t)${]} have found a global tendency hinting towards deceleration
of synaptic weights \citep[Fig. 2]{Statman2014_17}, in line with
the majority of cases presented in \prettyref{fig:acceleration_transition}(c)
being decelerating. However, the results of \citep[Fig. 2]{Statman2014_17}
show large fluctuations with single data points in opposition to the
global tendency. Our findings suggest that these fluctuations could
partly be accounted for when conditioning on the rate and the $C_{v}$
of the spike train. These could, for instance, be obtained using sliding
averages of a length scale on which the synaptic weight does not change
much. Specifically, our results suggest that in a region of low rate
and low $C_{v}$, accelerating weights are overrepresented. Moreover,
optogenetic stimulation of a neuron (roughly varying $\mu$) could
change a synapse from an accelerating to a decelerating regime.

\section{Dynamics of learning\label{sec:Dynamics-of-learning}}

In this section we study the network and learning scheme sketched
in \prettyref{fig:joint_model}(c,d) (introduced in Secs.~\ref{subsec:Network-model}
and \ref{subsec:Training-procedure-1}). The dynamics can now be viewed
on two time scales: First, on the discrete \emph{macroscopic} time
scale (pattern index $k$, going from one training session to the
next), the data that the network is exposed to is exchanged. Since
subsequent pattern pairs are uncorrelated, the sequence of matrices
$W_{k}$ after the $k$th training session including homeostasis is
a Markov chain (see also \citep{Amit1994_982}). Second, on the continuous
\emph{microscopic} time scale (within a training session), the present
pattern pair $(\bq_{k},\bp_{k})$ appears static, and the fast neural
and synaptic dynamics evolve as defined in \prettyref{sec:models}.
As we discuss in detail in this section, the system becomes stationary
on the macroscopic time scale while remaining non-equilibrium on the
microscopic time scale. The following analysis starts by studying
the storage of a memory, namely the association $\bq_{k}\rightarrow\bp_{k}$.
This storage occurs on the microscopic time scale by exposing the
system to $(\bq_{k},\bp_{k})$. The analysis proceeds by quantifying
the stationary regime on the macroscopic time scale. Assuming this
regime, we lastly study how memories are forgotten on the macroscopic
time scale.

\subsection{Synaptic dynamics on the microscopic time scale\label{subsec:Synaptic-dynamics-micro}}

Here, we characterize the change of the synaptic weight matrix $W(t)$
during training session $k$. The starting point is the homeostatically
scaled matrix from the previous session $W(0)=W_{k-1}$, where, as
mentioned in the discussion of \prettyref{eq:homeostasis-1}, each
row of $W(0)$ sums to $Mm_{0}$. The new pattern pair $(\bq_{k},\bp_{k})$
defines four synaptic populations, 
\begin{equation}
P_{ab}=\{W_{ij}:\,p_{k,i}=a,\,q_{k,j}=b\}.\label{eq:sets}
\end{equation}
These sets sort synapses by where they point to ($a=0$: to non-target
neurons, $a=1$: to target neurons) and where they come from ($b=0$:
from non-cue neurons, $b=1$: from cue neurons). Since $W(0)$ is
still uncorrelated to $(\bq_{k},\bp_{k})$, the weights in all populations
can be seen as randomly sampled from the set of weights in $W_{k-1}$.
We describe the synaptic dynamics through the population means and
variances
\begin{equation}
\hat{m}_{ab}=\frac{1}{|P_{ab}|}\sum_{w\in P_{ab}}w,\quad\hat{V}_{ab}=\frac{1}{|P_{ab}|}\sum_{w\in P_{ab}}(w-\hat{m}_{ab})^{2},
\end{equation}
where $|\cdot|$ here denotes the number of elements in the sets defined
in \prettyref{eq:sets}. Strictly speaking, $\hat{m}_{ab}$ and $\hat{V}_{ab}$
are stochastic processes that depend on the realizations of the Poissonian
and Gaussian processes fed into the network. Neglecting correlations,
according to the central limit theorem one would estimate their fluctuations
to scale as $1/\sqrt{|P_{ab}|}$, thus, because here populations are
assumed to be large, it is reasonable to assume self-averaging $\hat{m}_{ab}\approx\left\langle \hat{m}_{ab}\right\rangle \equiv m_{ab}$
and $\hat{V}_{ab}\approx\left\langle \hat{V}_{ab}\right\rangle \equiv V_{ab}$,
which strongly simplifies the analysis. The initial means are $m_{ab}(0)=m_{0}$
due to homeostasis, the initial variances $V_{ab}(0)=V_{0}$, where
here $V_{0}$ should be understood as a parameter; later we fix it
self-consistently to the stationary total variance of $W_{k\rightarrow\infty}$.

Our aim is to obtain $m_{ab}(t)$ and $V_{ab}(t)$ similarly to the
single synapse case in \prettyref{sec:Stochastic-dynamics-of}. To
this end, we first map the neurons in \prettyref{eq:brunel_net_E-1}
to instances of the single neuron in \prettyref{eq:lif}, who's input
is a mean current and Gaussian white noise with fixed intensity. Thus,
we need to calculate the mean and noise intensity of the input to
the neurons. The input from the Poisson layer $\sum_{j}W_{ij}\eta_{j}$
to the neurons in the target population ($a=1$) or in the non-target
population ($a=0$) is determined by the first- and second-order statistics
of Poisson processes
\begin{align}
\mu_{a}^{\text{cue}} & =M[m_{a1}f_{c}\nu_{\text{hi}}+m_{a0}(1-f_{c})\nu_{\text{lo}}]\nonumber \\
D_{a}^{\text{cue}} & =\frac{1}{2}M[(m_{a1}^{2}+V_{a1})f_{c}\nu_{\text{hi}}+(m_{a0}^{2}+V_{a0})(1-f_{c})\nu_{\text{lo}}].
\end{align}
The total input (including cue, recurrent network, and supervision)
to neurons in population $a$ is thus by Poisson approximation as
in \citep{Brunel00_183}
\begin{align}
\mu_{a}^{\text{tot}} & =\mu_{E}+JC_{E}r_{E}-gJC_{I}r_{I}+\mu_{a}^{\text{cue}}+\delta_{a1}J_{s}\nu_{s},\nonumber \\
D_{a}^{\text{tot}} & =D_{E}+\frac{1}{2}[J^{2}C_{E}r_{E}+(gJ)^{2}C_{I}r_{I}+D_{a}^{\text{cue}}+\delta_{a1}J_{s}^{2}\nu_{s}],\label{eq:tot_intput_a}
\end{align}
where $\delta_{a1}$ is the Kronecker symbol and $r_{E}$ and $r_{I}$
are the mean firing rates of the excitatory and the inhibitory neurons,
respectively. These are determined by mean-field theory along the
lines of \citep{Brunel00_183}, see \prettyref{appendix:MFT_recurrent}.

\begin{figure}
\includegraphics{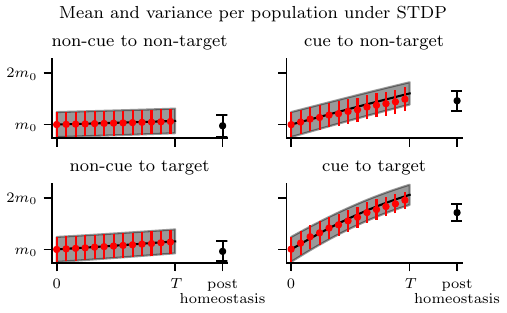}

\caption{\label{fig:Four-populations-training-session}Population dynamics
during a training session of length $T=50$. Mean $m_{ab}$ and standard
deviation $\sqrt{V_{ab}}$ of the four populations from \prettyref{eq:mVdynamics_4pops}
(black line and gray shading) and from simulations (red errorbars).
The separate rightmost errorbars schematically illustrate the effect
of homeostasis \prettyref{eq:homeostasis_theory} (we do not consider
an actual time for homeostasis). Parameters are \emph{neuron:} $\mu_{E}=0$,
$D_{E}=0.1$, $\mu_{I}=0.5$, $D_{I}=0.05$ \emph{network: }$N_{E}=4000$,
$N_{I}=1000$, $C_{E}=200$, $C_{I}=50$, $(J,g,h)=(0.01,5,2)$ \emph{input:
}$(f_{c},f_{s})=(0.05,0.1)$, $M=200$, $m_{0}=0.05$, $\nu_{\text{hi}}=1$,
$\nu_{\text{lo}}=0.1$, $\nu_{s}=64$, $J_{s}=1/80$, \emph{STDP:}
as in \prettyref{fig:joint_model}.}
\end{figure}

As one can see from the total effective input \prettyref{eq:tot_intput_a},
the postsynaptic neurons are, within the employed approximation, effectively
decoupled apart from their common dependence on the mean fields $m_{ab}$,
$V_{ab}$, $r_{E}$, and $r_{I}$, which we now determine self-consistently.
Thus, assuming knowledge of the numerical values of the mean fields,
the single-neuron statistics is readily determined and we can proceed
as for the single synapse in \prettyref{sec:Stochastic-dynamics-of}:
The drift- and diffusion coefficients of the four distinct synaptic
populations are, in analogy to Eqs.~\eqref{eq:D1} and \eqref{eq:D2},
respectively
\begin{align}
D_{ab}^{(1)}(w) & =(\Delta_{c}\tau_{c}-r_{ac}w\tau_{ac})\nu_{b}r_{a}(m,V)\nonumber \\
 & \;+\Delta_{c}\nu_{b}\left[w\alpha_{a}(i\tau_{c}^{-1},m,V)+\frac{1}{2}w^{2}\beta_{a}(i\tau_{c}^{-1},m,V)\right],\nonumber \\
D_{ab}^{(2)}(w) & =\frac{1}{4}r_{a}(m,V)\nu_{b}(\Delta_{c}^{2}\tau_{c}+r_{ac}^{2}w^{2}\tau_{ac}).\label{eq:D1D2_MFT}
\end{align}
Note that $D^{(1)}$ and $D^{(2)}$ depend parametrically on the population-wise
mean and variance of weights $m$ and $V$, since these impact the
rate and response of postsynaptic neurons. Lastly, we may close the
self-consistency by identifying $m_{ab}$ and $V_{ab}$ with the ensemble
average of synapses with drift- and diffusion coefficients \prettyref{eq:D1D2_MFT}.
Employing the method from \prettyref{subsec:solving_single_synapse}
leading to Eqs.~\eqref{eq:solution_mean_drift} and \eqref{eq:ode_variance},
this leads to 
\begin{align}
\dot{m}_{ab} & =D_{ab}^{(1)}(m_{ab};m,V)\nonumber \\
\dot{V}_{ab} & =2D_{ab}^{(1)\prime}(m_{ab};m,V)V_{ab}\nonumber \\
 & \;+\frac{r_{a}(m,V)\nu_{b}}{2}\left[\Delta_{c}^{2}\tau_{c}+r_{ac}^{2}\tau_{ac}\left(V_{ab}+m_{ab}^{2}\right)\right],\label{eq:mVdynamics_4pops}
\end{align}
where the drift coefficient depends on $m$ and $V$ both as its argument
and through the parametric dependence mentioned above. To summarize,
the dynamics of the means and variances of the four populations in
training session $k$ is given in terms of eight coupled differential
equations \prettyref{eq:mVdynamics_4pops} which we integrate numerically.
In \prettyref{fig:Four-populations-training-session}, we show $m_{ab}$
and $\sqrt{V_{ab}}$ as functions of time. The mean weight of all
four populations grows, reflecting that their reversal points are
above $m_{0}$.  

\subsection{Stationarity on the macroscopic time scale}

After evolving with STDP for the training time $T$, the synapses
undergo homeostasis: the weights are scaled so as to fix the mean
weight in $W$ per row to $m_{0}$. This requires rescaling of $W_{ij}\in P_{ab}$
with 
\begin{equation}
\gamma_{a}=\frac{m_{0}}{f_{c}m_{a1}(T)+(1-f_{c})m_{a0}(T)}.\label{eq:homeo_scalor}
\end{equation}
Thus, after the $k$th training session and the subsequent homeo\-static
process, the population means and variances of $W_{k}$ are 
\begin{equation}
\gamma_{a}m_{ab}(T)\,\,\text{and}\,\,\gamma_{a}^{2}V_{ab}(T),\label{eq:homeostasis_theory}
\end{equation}
respectively, these are the separate errorbars in \prettyref{fig:Four-populations-training-session}.

While the total average of $W_{k}$, 
\begin{equation}
N_{E}^{-1}M^{-1}\sum_{ij}W_{k,ij}=m_{0},
\end{equation}
 is fixed by construction, the same is not true for the total variance
\begin{equation}
V_{k}=N_{E}^{-1}M^{-1}\sum_{ij}(W_{k,ij}-m_{0})^{2}.
\end{equation}
However, $V_{k}$ approaches a stationary value due to the interplay
of STDP and homeostasis: Assuming the total variance in the beginning
of the $k$th training session was $V_{\ast}$, this fixes the initial
condition of all four variances in \prettyref{eq:mVdynamics_4pops}
to $V_{\ast}$. The four means and variances after STDP for time $T$
and subsequent homeostasis, \prettyref{eq:homeostasis_theory}, thus
parametrically depend on $V_{\ast}$. Therefore, the marginal variance
of weights in $W$ after the $k$th STDP session and homeostasis,
$V_{k}$, depends on $V_{\ast}$ as well. It can be expressed in terms
of the results of integrating \prettyref{eq:mVdynamics_4pops} as
\begin{align}
V_{k}(V_{\ast}) & =\sum_{ab}^{\ast}[\gamma_{a}^{2}V_{ab}(T)+(\gamma_{a}m_{ab}(T)-m_{0})^{2}],
\end{align}
where we defined the weighted population sum
\begin{equation}
\sum_{ab}^{\ast}X_{ab}\equiv\begin{pmatrix}1-f_{s}, & f_{s}\end{pmatrix}\begin{pmatrix}X_{00} & X_{01}\\
X_{10} & X_{11}
\end{pmatrix}\begin{pmatrix}1-f_{c}\\
f_{c}
\end{pmatrix}.
\end{equation}
The asymptotic variance must be stationary
\begin{equation}
V_{\ast}=V_{k}(V_{\ast}).\label{eq:stationary_variance}
\end{equation}
We solve \prettyref{eq:stationary_variance} with a bisection algorithm.
As tested with simulations {[}see \prettyref{fig:dynamics_macroscopic}(b){]},
the variance indeed approaches the result of \prettyref{eq:stationary_variance}
to satisfying accuracy; note that the variances shown in \prettyref{fig:dynamics_macroscopic}(b)
are variances of subsets of synapses, all of which converge to $V_{\ast}$.

\subsection{Forgetting on the macroscopic time scale}

Here, we take the final steps from neuron- and synapse models to a
key property of memory--the memory capacity. To this end, we compute
the degradation of a memory trace from which we estimate the fraction
of correctly activated neurons in an attempted recall.

\subsubsection{Trace degradation}

We assume that the system is in the macroscopically stationary state
identified above. After the training of a specific pattern pair,
say $(\bq_{0},\bp_{0})$, the weights from $\bq_{0}$ to $\bp_{0}$
have mean $K_{11}^{(0)}=\gamma_{1}m_{11}(T)$ and variance $G_{11}^{(0)}=\gamma_{1}^{2}V_{11}(T)$.
Analogously, the four distinct populations have mean and variance
\begin{align}
K^{(0)} & =\begin{pmatrix}\gamma_{0}m_{00}(T) & \gamma_{0}m_{01}(T)\\
\gamma_{1}m_{10}(T) & \gamma_{1}m_{11}(T)
\end{pmatrix}\\
\text{and}\,\,G^{(0)} & =\begin{pmatrix}\gamma_{0}^{2}V_{00}(T) & \gamma_{0}^{2}V_{01}(T)\\
\gamma_{1}^{2}V_{10}(T) & \gamma_{1}^{2}V_{11}(T)
\end{pmatrix}.
\end{align}
For a successful storage of the association $\bq_{0}\rightarrow\bp_{0}$
it is required that cue-to-target synapses are significantly stronger
than other synapses, i.e., that $K_{11}^{(0)}$ (the mean cue-to-target
weight) exceeds the other means beyond the uncertainty captured by
the variances $G^{(0)}$. This will be made rigorous below in \prettyref{subsec:Recall}.
Here, we ask how this elevation of synapses from $\bq_{0}$ to $\bp_{0}$
(the reference association) decays if additional (competing) associations
are trained. For instance, we track the mean $K_{11}^{(k)}$ of synapses
from $\bq_{0}$ to $\bp_{0}$ due to the storage of $k$ subsequent
training phases. While the matrices $K^{(k)}$ and $G^{(k)}$ refer
to means and variances of synapses sorted by their role in the reference
association, training the $k$th competing association $(\bq_{k},\bp_{k})$
induces changes in all four reference populations. For instance,
among the $f_{s}N_{E}\times f_{c}M$ synapses from $\bq_{0}$ to $\bp_{0}$,
roughly $f_{c}^{2}f_{s}^{2}MN_{E}$ synapses are also cue-to-target
synapses in association $k$, and $f_{c}^{2}f_{s}(1-f_{s})MN_{E}$
synapses are cue-to-nontarget in association $k$. More generally,
within each reference population $(a,b)$, four subpopulations $(a^{\prime},b^{\prime})$
evolve differently. Assuming large enough neuron numbers, each subpopulation
covers a fraction $[\delta_{a^{\prime}0}(1-f_{s})+\delta_{a^{\prime}1}f_{s}][\delta_{b^{\prime}0}(1-f_{c})+\delta_{b^{\prime}1}f_{c}]$
of the surrounding block $(a,b)$ each of which consists of $[\delta_{a0}(1-f_{s})+\delta_{a1}f_{s}][\delta_{b0}(1-f_{c})+\delta_{b1}f_{c}]MN_{E}$
synapses. Thus, the mean and variance of each block $(a,b)$ changes
from $K_{ab}^{(k-1)},G_{ab}^{(k-1)}$ to $K_{ab}^{(k)},G_{ab}^{(k)}$
due to the drift and diffusion of its subpopulations $(a^{\prime},b^{\prime})$
all of which are initialized at mean $m_{a^{\prime}b^{\prime}}(0)=K_{ab}^{(k-1)}$
and variance $V_{a^{\prime}b^{\prime}}(0)=G_{ab}^{(k-1)}$. Strictly
speaking, the problem explodes at this point since $4\times4$ synaptic
populations together determine their stochastic dynamics through the
interaction with the heterogeneous recurrent network with different
initial statistics at each training step. To study the change $K_{ab}^{(k-1)}\to K_{ab}^{(k)}$
we compute the effective shift $m_{a^{\prime}b^{\prime}}(T)|_{m_{a^{\prime}b^{\prime}}(0)}-m_{a^{\prime}b^{\prime}}(0)$
(where the subscript denotes the initial value) in each subpopulation.
This shift depends only weakly on the initial value $m_{a^{\prime}b^{\prime}}(0)$
such that the precise initial value does not matter much and can be
set to the mean value $m_{0}$
\[
m_{a^{\prime}b^{\prime}}(T)\big|_{K_{ab}^{(k-1)}}-K_{ab}^{(k-1)}\approx m_{a^{\prime}b^{\prime}}(T)\big|_{m_{0}}-m_{0}.
\]
We can thus express the effect of one training session on $K^{(k)}$
as a simple map
\begin{align}
K_{ab}^{(k)} & =\sum_{a^{\prime}b^{\prime}}^{\ast}\gamma_{a^{\prime}}\left(K_{ab}^{(k-1)}+[m_{a^{\prime}b^{\prime}}(T)-m_{0}]\right).\label{eq:trace_degrad_iter}
\end{align}
Defining 
\begin{align}
\phi & =1-\sum_{a=0}^{1}[\delta_{a0}(1-f_{s})+\delta_{a1}f_{s}]\gamma_{a},\nonumber \\
c & =\sum_{ab}^{\ast}\gamma_{a}[m_{ab}(T)-m_{0}],\label{eq:trace_degrad_coeffs}
\end{align}
we can recast \prettyref{eq:trace_degrad_iter} into
\begin{equation}
K_{ab}^{(k)}=(1-\phi)K_{ab}^{(k-1)}+c.
\end{equation}
This recursive equation has the solution
\begin{equation}
K_{ab}^{(k)}=(1-\phi)^{k}K_{ab}^{(0)}+\phi^{-1}c[1-(1-\phi)^{k}]\label{eq:trace_degrad_sol}
\end{equation}
as one can check by insertion. Thus, for small $\phi$, the elevated
weights relax with rate $\phi$; this relaxation rate can be fully
expressed through the rescaling factors $\gamma_{a}$. If $\gamma_{a}=1$
(i.e., no rescaling necessary to maintain $m_{0}$), the trace would
not be degraded at all. For $\gamma_{a}<1$ (compensating for STDP
that on average potentiates), it is rather homeostatic downscaling
and less overwriting by new memories that limits the lifetime of a
memory. Asymptotically, all $K_{ab}^{(k)}$ relax to the imposed population
mean $m_{0}=\phi^{-1}c$, as one can check by inserting \prettyref{eq:homeo_scalor}
into \prettyref{eq:trace_degrad_coeffs}.

An analogous argument for the variances leads to 
\begin{align}
G_{ab}^{(k)} & =\sum_{a^{\prime}b^{\prime}}^{\ast}\gamma_{a^{\prime}}^{2}\Big[G_{ab}^{(k-1)}+V_{a^{\prime}b^{\prime}}(T)-V_{\ast}\nonumber \\
 & \;+\left(K_{ab}^{(k-1)}+m_{a^{\prime}b^{\prime}}(T)-m_{0}\right)^{2}\Big]-K_{ab}^{(k)2}\label{eq:var_degrad_iter}
\end{align}
for which we do not know an explicit solution. Still, \prettyref{eq:var_degrad_iter}
can be evaluated numerically. $K^{(k)}$ and $G^{(k)}$ are shown
and compared to simulations in \prettyref{fig:dynamics_macroscopic}(a,b);
within a few hundred patterns, $K^{(k)}$ and $G^{(k)}$ decay to
their equilibrium values $m_{0}$ and $V_{\ast}$, respectively.

\subsubsection{Recall\label{subsec:Recall}}

To get a meaningful measure of memory capacity, we investigate how
well target 0 can be recalled when cue 0 is presented after $k$ subsequent
training sessions. For simplicity, we only consider the case $\nu_{\text{lo}}=0$,
i.e., where non-cue Poisson processes are silent. We follow the approach
of \citep{Auer2025_preprint} and evaluate the dendritic sums: For
each post-synaptic neuron, we compute its summed synaptic weight stemming
from cue 0 neurons 
\begin{equation}
s_{i}=\sum_{j}W_{ij}q_{0,j}.\label{eq:summed_weight}
\end{equation}
The summed weight reflects how strongly each neuron is driven by the
cue 0. If the $f_{s}N_{E}$ most strongly driven neurons are the target
pattern $\bp_{0}$, the association is perfectly recalled. We assume
that a certain amount of errors can be corrected by a downstream mechanism,
e.g., by an attractor network or a perceptron. We denote neurons among
the most strongly driven neurons that are indeed in $\bp_{0}$ as
the correctly activated neurons. Next, we derive the fraction $a_{k}$
of correctly activated neurons in an attempted recall $k$ training
sessions after retrieval. The fraction $a_{k}$ may be interpreted
as the accuracy of an association. From the accuracy, we define the
memory capacity as
\begin{equation}
c=\text{argmin}_{k}(|a_{k}-0.5|),\label{eq:capacity}
\end{equation}
which is the number of patterns $c$ after which the accuracy is below
$50\%$, $a_{k=c}\leq0.5$, i.e., we have more false positives than
correctly activated neurons. To derive the capacity we have to study
the overlap between summed-weight distributions into target and non-target
neurons, respectively: If this overlap vanishes, target neurons receive
significantly stronger input than non-target neurons and can thus
be clearly distinguished. If the distributions strongly overlap, it
is hard to distinguish target and non-target neurons by their activity.
These considerations are detailed next.

\begin{figure}[t]
\includegraphics{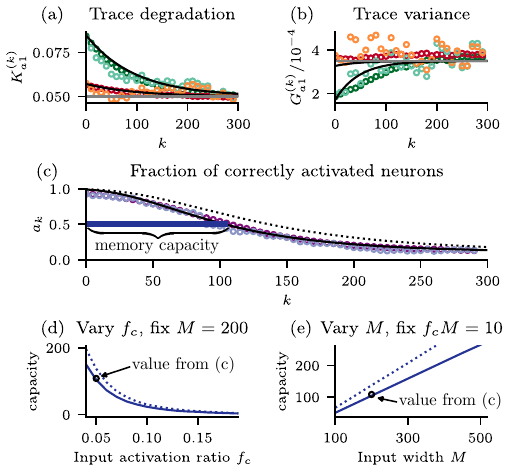}

\caption{\label{fig:dynamics_macroscopic}Dynamics of memory traces and memory
capacity. (a) Mean synaptic strength from cue to (non-)target neurons
(with respect to association 0) after $k$ subsequently stored associations.
Simulation results denoted by circles, green for cue-to-target and
red for cue-to-non-target, dark colors are averages over $100$ realizations,
light colors are single realizations. Theory \prettyref{eq:trace_degrad_sol}
(black lines) and $m_{0}$ (gray line). (b) Variance of synaptic strength
from cue to (non-)\-target. Colors as in (a), theory (black) from
\prettyref{eq:var_degrad_iter} and stationary variance (gray) from
\prettyref{eq:stationary_variance}. (c) Fraction of correctly activated
neurons on average over 100 realizations (dark purple circles), for
a single realization (light circles) and theory \prettyref{eq:frac_correct_hits}
based on the full drift (black solid line) and neglecting cross-correlations
(dotted line). (d,e) Memory capacity \prettyref{eq:capacity} as a
function of input sparseness $f_{c}$ (d) and the input width $M$
(e) with (solid) and without (dotted) cross-correlations. Parameters
as in \prettyref{fig:Four-populations-training-session}, but $\nu_{\text{lo}}=0$.}
\end{figure}

Assuming weak correlations between elements of $W$ we can use the
central limit theorem to study \prettyref{eq:summed_weight}; the
summed weight to target neurons ($a=1$) and non-target neurons ($a=0$)
is respectively a Gaussian random variable
\begin{equation}
p_{a}(s_{i})=\cN(s_{i}|f_{c}M\,K_{a1}^{(k)},f_{c}M\,G_{a1}^{(k)}),\label{eq:clt_summed_weight}
\end{equation}
where the mean and variance are given by the block statistics \prettyref{eq:trace_degrad_sol}
and \prettyref{eq:var_degrad_iter}. Marginalizing over all post-synaptic
neurons (i.e., forgetting their identity as target or non-target neuron
with fractions $f_{s}$ and $1-f_{s}$, respectively), the distribution
of summed weight $s$ is 
\begin{align}
p_{\text{marg}}(s)= & f_{s}\cN(s|f_{c}M\,K_{11}^{(k)},f_{c}M\,G_{11}^{(k)})\nonumber \\
 & +(1-f_{s})\cN(s|f_{c}M\,K_{01}^{(k)},f_{c}M\,G_{01}^{(k)}).\label{eq:summed_weight_dist}
\end{align}
The $f_{s}N_{E}$ most active neurons have $s\geq s_{\ast}$, where
$s_{\ast}$ is the $1-f_{s}$th quantile of $p_{\text{marg}}$, i.e.,
\begin{equation}
\int_{s_{\ast}}^{\infty}ds\,p_{\text{marg}}(s)=f_{s}.
\end{equation}
This integral can be evaluated to 
\begin{equation}
\frac{f_{s}}{2}\text{erfc}\left(\frac{s_{\ast}-f_{c}M\,K_{11}^{(k)}}{\sqrt{2f_{c}MG_{11}^{(k)}}}\right)+\frac{1-f_{s}}{2}\text{erfc}\left(\frac{s_{\ast}-f_{c}M\,K_{01}^{(k)}}{\sqrt{2f_{c}MG_{01}^{(k)}}}\right)=f_{s},\label{eq:quantile_erfc}
\end{equation}
where $\text{erfc}$ is the complementary error function. We solve
\prettyref{eq:quantile_erfc} numerically. Finally, the fraction of
correctly activated neurons is the fraction of probability mass of
\prettyref{eq:summed_weight_dist} above $s_{\ast}$ that is due to
target neurons 
\begin{align}
a_{k} & =\int_{s_{\ast}}^{\infty}ds\,\cN(s|f_{c}MK_{11}^{(k)},f_{c}MG_{11}^{(k)})\nonumber \\
 & =\frac{1}{2}\text{erfc}\left(\frac{s_{\ast}-f_{c}M\,K_{11}^{(k)}}{\sqrt{2f_{c}MG_{11}^{(k)}}}\right).\label{eq:frac_correct_hits}
\end{align}
\prettyref{eq:frac_correct_hits} is shown in \prettyref{fig:dynamics_macroscopic}(c).
From \prettyref{eq:frac_correct_hits} one can compute the memory
capacity. As well known from Hopfield-like networks, training with
sparse patterns is less detrimental to previous memory \citep{Palm2013_171};
this observation remains valid for the spike-coding setup, as shown
in \prettyref{fig:dynamics_macroscopic}(d). Furthermore, we observe
in \prettyref{fig:dynamics_macroscopic}(c-e) that rate-based approximations
systematically and substantially overestimate the memory capacity.
Our theory that incorporates spike-time--resolving cross-correlations
correctly predicts memory capacities much more faithfully. Lastly,
memory capacity grows linearly with the width $M$ of the input layer
{[}\prettyref{fig:dynamics_macroscopic}(e){]}.

\subsection{Tradeoff between early accuracy and capacity}

Neural systems are often forced to trade one benefit for another,
for instance, speed-accuracy tradeoffs have been reported both experimentally
and theoretically, e.g. in decision making \citep{Heitz2012_628}
and in memory consolidation \citep{Bhasin2024_e2406010121}. We here
discuss a tradeoff between early accuracy and memory lifetime.

Naively, one could assume that the higher the early accuracy (here
defined as the fraction of correctly activated neurons right after
training, $a_{0}$), the higher the memory capacity. This is not always
the case. In \prettyref{fig:Tradeoffs}(a), we vary different parameters
of the plasticity rule and the training scheme and find that the optimal
capacity does not coincide with the maximum early accuracy, i.e.,
the maximal capacity is not at $a_{0}=1$. Specifically, when varying
the amplitude of potentiation $\Delta_{c}$, the time scale of potentiation
$\tau_{c}$, or the length of a training session $T$, the capacity
is optimal for $a_{0}<1$. It thus seems as if a system must decide
to either optimize capacity \emph{or} early accuracy.

A way to avoid the tradeoff altogether is to simultaneously change
both the STDP amplitude $\kappa(\tau)\to\rho\kappa(\tau)$ and the
training time $T\to T/\rho$. The scaling parameter $\rho$ determines
the learning speed but keeps the mean effect of a training session
on cue-to-target synapses constant. The dilemma is in fact resolved:
lowering the speed $\rho$ increases both capacity and early accuracy
thus evading the capacity-accuracy tradeoff, see \prettyref{fig:Tradeoffs}(b).
If speed itself is considered a desired benefit, this means we have
a speed-accuracy-capacity triple tradeoff and maximizing capacity
and accuracy comes at the expense of a low learning speed.

\subsection{Optimal exposure time}

\begin{figure}
\includegraphics{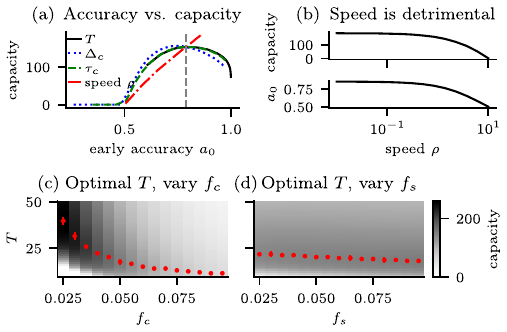}

\caption{\label{fig:Tradeoffs}How to get high capacity? (a) Memory capacity
against early accuracy $a_{0}$ when varying $T$ (black solid), $\Delta_{c}$
(blue dotted), $\tau_{c}$ (green dashed), and speed (red dash-dotted).
In order to understand how the parameters are varied, note that $a_{0}(T)$,
$a_{0}(\Delta_{c})$, and $a_{0}(\tau_{c})$ are all increasing functions
of their argument. Speed here means upscaling of $\kappa(\tau)$ while
downscaling $T$. (b) Capacity and early accuracy saturate for low
speed and decrease with large speed. In (a) and (b), unless explicitly
varied, $T=17$, which optimizes the capacity at parameters from \prettyref{fig:dynamics_macroscopic}.
(c,d) Memory capacity gray-coded for various $f_{c}$ (c) or $f_{s}$
(d) and $T$. The red bars show the regime of optimal capacity for
a given sparseness $f_{c}$ or $f_{s}$. Parameters that are not varied
are as in \prettyref{fig:dynamics_macroscopic}.}
\end{figure}

The heteroassociative network studied in this paper only captures
a single stage in memory formation, which likely involves several
stages and parallel pathways. Yet, assuming that the performance of
single stages is related to the performance of the whole system, we
can even derive qualitative predictions about behavior: The duration
$T$ of a training session can likely be behaviorally controlled by
changing the time a proband is exposed to two stimuli they are supposed
to associate (e.g., a sound and an image). Is there an optimal duration
of exposure or is longer always better? To answer this question we
follow the idea of \citep{Auer2025_preprint} and determine the time
$T$ (corresponding to the transition probability in \citep{Auer2025_preprint})
that optimizes capacity. As shown in \prettyref{fig:Tradeoffs}(c,d),
the memory capacity is optimized at a finite exposure time. The quantitative
value of the optimal time is quite sensitive to the presynaptic sparseness
(c), yet quite insensitive to the postsynaptic sparseness (d).

\subsection{Two complementary roles of homeostasis}

Homeostasis, i.e., the slow scaling of weights (during the breaks
between learning sessions) modeled by \prettyref{eq:homeostasis-1}
is both responsible for forgetting and a requirement for learning.
The responsibility for forgetting is apparent from \prettyref{eq:trace_degrad_sol}:
The elevated mean weight of cue-to-target synapses degrades with rate
$\phi$, an important ingredient of which are the homeostatic scaling
factors $\gamma_{a}$, see \prettyref{eq:trace_degrad_coeffs}. However,
as discussed now, without homeostasis, in the long run, new associations
cannot even be learned.

\begin{figure}
\includegraphics{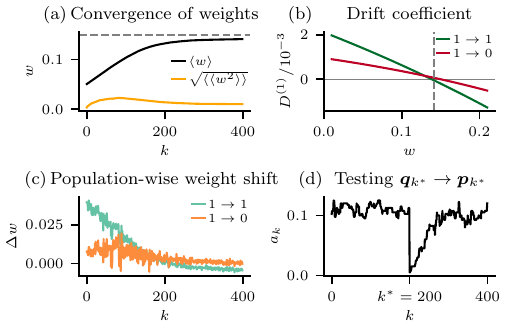}

\caption{\label{fig:Learning-without-homeostasis.}Training without homeostasis.
(a) Transient population mean and standard deviation of the synaptic
weights over several training steps $k$. The dashed line is the weighted
sum $f_{s}w_{11}+(1-f_{s})w_{01}$ of the zeros $w_{11}$ ($w_{01}$)
of the drift coefficient of cue-to-target (cue-to-non-target) synapses,
respectively, see (b). (b) Drift coefficient \prettyref{eq:D1D2_MFT}
of cue-to-target (green) and cue-to-non-target (red) synapses assuming
the final values in (a) for the statistics of the recurrent network.
Horizontal line at zero. Vertical dashed line at asymptotic mean weight
from (a). (c) Difference of synaptic weight before and after training
step $k$, averaged over cue-to-target synapses in pattern pair $k$
(turquoise) and over cue-to-non-target synapses in pattern pair $k$
(orange). (d) Fraction of correctly activated neurons when attempting
to retrieve an association that is trained at step $k^{\ast}=200$.
Parameters as in \prettyref{fig:dynamics_macroscopic} but with homeostasis
deactivated.}
\end{figure}

If we omit the homeostatic scaling, then the overall mean synaptic
weight $m_{0}$ is not enforced anymore. Instead, the mean weight
converges to a stable fixed point {[}black line in \prettyref{fig:Learning-without-homeostasis.}(a){]}
with the standard deviation (orange line) being bounded, i.e., the
weights do not diverge. The fixed point of the mean weight is around
the zeros of the drift coefficients of cue-to-target and cue-to-non-target
synapses, \prettyref{fig:Learning-without-homeostasis.}(b). In the
stationary regime, is it still possible to train patterns? To train
pattern $k$, we require that the synapses from $\bq_{k}$ to $\bp_{k}$
grow significantly, i.e., that the difference of cue-to-target synaptic
weight $\Delta w$ over the $k$th training session is positive and
large. Specifically, the $\Delta w$ of cue-to-non-target synapses
must be smaller. These two competing shifts are compared in \prettyref{fig:Learning-without-homeostasis.}(c)
and display a surprising behavior: at low $k$ where we start the
weights below the fixed point, training works; however, towards the
stationary regime (large $k$), i.e., around the fixed point, the
picture is reversed and we observe a kind of anti-learning. This is
demonstrated by tracking the retrievability of an association, namely
for the pattern pair applied at session $k^{\ast}=200$ (close to
the stationary regime). Indeed, after training this pattern pair,
it is \emph{less likely than chance} to have correctly activated neurons
in recall, reflected by the downstroke in \prettyref{fig:Learning-without-homeostasis.}(d).

We conclude that while the STDP rule by \citep{Bi1998_10472} is inherently
stable, it does not allow for Hebbian association learning without
a homeostasis mechanism shifting it far from the STDP rule's inherent
stable fixed point. Moreover, we hypothesize that deactivation of
homeostatic plasticity not only decreases the memory capacity, but
can even lead to anti-learning.

\section{Discussion}

In this paper we analytically described the stochastic process of
the weight of synapses endowed with STDP and driven by stochastic
spike trains of pre- and postsynaptic neurons. Specifically, we derived
a concise description in form of a Langevin equation that captures
the first two Kramers-Moyal coefficients of the true jump process.
From this description we computed the dynamics of the mean and variance
of an ensemble of synaptic weights. We next studied a training setup
where hetero-associations are stored into a feed-forward matrix of
synapses endowed with STDP. Through a mean-field theory, we mapped
this setup to the single-synapse case above, which led to a quantitative
description of the memory lifetime. Training and retrieval in this
setting are successful (with memory capacity scaling linearly with
the input width) for a range of parameters, thus the procedure is
robust. We made qualitative predictions about the drift and diffusion
coefficients of single synapses, about a tradeoff between early accuracy
and memory capacity, and about the role of homeostatic plasticity
in memory.

We included the effect of pre-post cross-correlations on the synaptic
dynamics through an exact link between the cross-correlation and the
neuron's response functions \citep{Stubenrauch2024_041047}. We found
that both for the single-synapse process and for the network process,
cross-correlations have a significant impact if synaptic weights are
strong enough. The discrepancy between the full solution and a rate-based
approximation is particularly prominent at sparse input patterns.
Thus, especially in sparse-coding situations, the theory developed
here is an important advancement over rate-based approaches.

Two research perspectives on memory and learning have been united
in this paper. On the one hand, in the context of STDP, research often
focuses on the transient (i.e., non-stationary) dynamics of ensembles
of synaptic weights \citep{Kempter1999_4514,vanRossum2000_8821,Burkitt2004_940,Meffin_2006_041911,Gilson2009_102,Fremaux2016_85,Ocker2018_951,Akil2021_23,Sosis2024_bioRxiv}.
These dynamics can, e.g., be expressed by continuous-time differential
equations of moments of the synaptic weights. The process described
in our theory corresponds to this type of dynamics during each training
session, which we referred to above as the microscopic time scale.
On the other hand, memory is often described as a discrete update
process of a weight matrix happening at a macroscopic timescale \citep{Hopfield1982_2558,Nadal1986_535,Buckingham1992,Amit1994_982,Tyulmankov2021_22258,Iatropoulos2022_35338,Auer2025_preprint}.
Thus, with each discrete time step, a new association is stored such
that it can partially overwrite and interfere with previous associations
(termed palimpsest \citep{Nadal1986_535}). If such a learning scheme
incorporates homeostasis, this discrete-time process approaches stationary
dynamics. In our process, this second perspective is covered, too:
when subsampling the process once after each training session including
homeostasis, the weight-matrices $W_{k}$ approach a stationary Markov
chain.

Interestingly, we found that the degradation of the memory cue must
first be attributed to homeostasis {[}see \prettyref{eq:trace_degrad_coeffs}
and \prettyref{eq:trace_degrad_sol}{]} and thus only indirectly occurs
due to the storage of new memory. Effectively, the memory degradation
with homeostatic plasticity is thus relaxational as opposed to the
palimpsest-like forgetting due to overwriting \citep{Nadal1986_535};
apparently the representations are large enough for overwriting effects
to average out for many training steps and the delimiting factor is
homeostatic relaxation. It would be interesting to clarify experimentally
under which circumstances forgetting is rather relaxational or palimpsest-like.

The hetero-associative setting with training induced by exposure,
\prettyref{fig:joint_model}(c), can model memory dynamics on multiple
stages. Assuming that the cue and target patterns are sensory representations,
our setting models initial retrieval of associations. Additionally,
the cue and/ or target patterns could be set by a different synaptic
pathway, a teacher. In the context of memory consolidation, the teacher
may be a synaptic pathway via the hippocampus which consolidates memory
by transferring the association to more stable pathways, as discussed
in \citep{Remme2021_37}.

The success of the approach at hand opens a vast set of intriguing
directions: Previous theory on recurrent plasticity (e.g., Refs. \citep{Ocker2018_951,Kossio2021})
should be revisited from the present point of view to justify approximations
and find corrections. In such recurrent settings, one might need to
go beyond the Poissonian-input approximation by applying the results
on colored shot noise in \citep{Stubenrauch2024_041047} and on self-consistent
power spectra in \citep{Dummer14}. When adding recurrent plasticity
to the post-synaptic population in our setup, attracting rate states
can arise which certainly impact memory, especially if one considers
correlated training patterns.

It has been experimentally reported that synaptic weights fluctuate
even in the absence of neural activity \citep{Yasumatsu2008_13608,Statman2014_17};
consequently, one could also study an extension of our model in which
the intrinsic synaptic noise is taken into account. While the change
of sign of the drift coefficient {[}see \prettyref{fig:single_synapse_full}(b){]}
has been experimentally reported, e.g. in \citep{Statman2014_17},
the non-monotonic dependence of the diffusion coefficient {[}see \prettyref{fig:single_synapse_full}(c){]}
has to our knowledge not been reported and should be investigated
experimentally.

Another interesting direction concerns the recent finding that different
transfer functions in multi-layer perceptrons give rise to qualitatively
different representation (or coding) schemes in the feature layer
\citep{vanMeegen2025_3354}; correspondingly, it would be quite interesting
to study multi-stage feed-forward and locally recurrent networks of
spiking neurons under STDP. When exposing the input and readout layer
to data, as in the present paper, representations in the hidden layer
will arise. It would be interesting to study the statistics of these
representations.

Lastly, while for rate-based neural networks the joint neural and
synaptic dynamics have been comprehensively described \citep{Clark2024_021001},
a corresponding theory for spiking (integrate-and-fire) neurons is
still missing; the approach presented here is a step in that direction.
\begin{acknowledgments}
We acknowledge helpful discussions with Holger Kantz and Moritz Helias.
This work has been funded by the Deutsche Forschungsgemeinschaft (DFG,
German Research Foundation), SFB1315, project-ID 327654276 to BL and
RK.
\end{acknowledgments}

\appendix

\section{Numerical methods\label{appendix:Numerical-methods}}

\subsection{Implementation}

Simulations of synapses and neurons are implemented in cython \citep{behnel2010cython}.
For the LIF neurons, \prettyref{eq:lif}, membrane voltages are integrated
with the Euler-Maruyama scheme. Numerical Kramers-Moyal coefficients
in \prettyref{fig:single_synapse_full} are obtained from simulations
that are initialized at $w(0)$ uniformly distributed in $[0,0.31]$
and thermalizing for a period of $T_{\text{warm}}=50$. In the network,
Eqs. \eqref{eq:brunel_net_E-1} and \eqref{eq:brunel_net_I-1}, spike
detection in one time step results in voltage reset and spike delivery
in the same time step. The synaptic model is simulated by integrating
\prettyref{eq:model_exp_local} with the Euler scheme. Additionally,
whenever the weight is set to a negative value, it is clipped to zero.
For the experimentally inspired parameters used in the present paper,
this almost never happens. For both \prettyref{eq:lif} and \prettyref{eq:model_exp_local},
the Euler time step is $\Delta t=10^{-4}\tau_{m}$.

Equations \eqref{eq:solution_mean_drift} and \eqref{eq:mVdynamics_4pops}
are integrated numerically using a Runge-Kutta scheme of order 5(4).

\subsection{Runtime}

To exemplify the runtime needed for evaluation of the theory and the
simulations, respectively, we here comment on the production of \prettyref{fig:dynamics_macroscopic}.
On an AMD Ryzen 5 PRO 4650U, the evaluation of all theoretical lines
took $\approx18.4\,\text{s}$ and the simulation took per realization
$\approx4.5\,\text{h}$. The 100 realizations used in \prettyref{fig:dynamics_macroscopic}
have been performed in parallel on a high-performance computing cluster.

\section{Expressing synaptic drift by cross-correlations\label{appendix:drift_by_cross_corr}}

Here, we provide detail of the steps from the synapse model \prettyref{eq:model_general}
to the synaptic drift in terms of cross-correlations \prettyref{eq:force}.
The drift coefficient is defined by \prettyref{eq:drift_def.}. As
opposed to the diffusion coefficient, the limit in \prettyref{eq:drift_def.}
can be written as a time derivative, since the expression in the expectation
value is linear in the trajectory $w_{\text{traj}}$. Thus, the model
can be directly plugged in such that 
\begin{align}
D^{(1)}(w) & =\int_{-\infty}^{t}dt^{\prime}\,\left\langle \kappa[t-t^{\prime},w_{\text{traj}}(t)]\eta(t^{\prime})x(t)\right\rangle \nonumber \\
 & +\int_{-\infty}^{t}dt^{\prime}\,\left\langle \kappa[t^{\prime}-t,w_{\text{traj}}(t)]\eta(t)x(t^{\prime})\right\rangle ,
\end{align}
where the average is over trajectories of the weight $w_{\text{traj}}$
under the condition that $w_{\text{traj}}(t)=w$. Due to this condition,
the multiplicativity of the synaptic depression (i.e., the dependence
of $\kappa$ on $w_{\text{traj}}$) does not complicate the analysis,
as
\begin{equation}
\kappa[t^{\prime}-t,w_{\text{traj}}(t)]\equiv\kappa(t^{\prime}-t,w).
\end{equation}
Thus, under the expectation value we can join the two integrals in
\prettyref{eq:model_general}
\begin{align}
D^{(1)}(w) & =\int_{-\infty}^{\infty}d\tau\kappa(\tau,w)\Big[\Theta(\tau)\left\langle x(t)\eta(t-\tau)\right\rangle \nonumber \\
 & +\Theta(-\tau)\left\langle x(t+\tau)\eta(t)\right\rangle \Big].\label{eq:almost_drift_by_cc}
\end{align}
The final step to \prettyref{eq:force} requires to assume stationarity
of the process $x$, since then the term in the square brackets of
\prettyref{eq:almost_drift_by_cc} becomes $\left\langle x\right\rangle \left\langle \eta\right\rangle +C_{x\eta}(\tau)$.
Since the synaptic weight $w$ may drift, $x$ is strictly speaking
non-stationary. However, on the small duration covered by the STDP
window (here, $\approx40\,\text{ms}$), the process appears stationary
since changes of the weights (and thus of the moments of $x$) are
slow for the biophysically motivated parameters used. Additionally,
since transient times of the LIF neuron are typically quick (only
a few inter-spike intervals if the coefficient of variation is big
enough \citep{Lindner02}), these stationary statistics may be evaluated
at the instantaneous value of $w$.

\section{Response functions of the LIF neuron\label{appendix:Response_LIF}}

The susceptibilities required in \prettyref{eq:D1} are the susceptibility
of the LIF neuron \prettyref{eq:lif} to mean- and noise-intensity-modulations,
respectively. These have been derived using Fokker-Planck theory in
\citep{Lindner2001_2937},
\begin{align}
\alpha(\Omega) & =\frac{ri\Omega/\sqrt{D_{\text{DA}}}}{i\Omega-1}\frac{\cD_{i\Omega-1}\left(\frac{\mu_{\text{DA}}-v_{\text{t}}}{\sqrt{D_{\text{DA}}}}\right)-e^{\Delta}\cD_{i\Omega-1}\left(\frac{\mu_{\text{DA}}-v_{\text{r}}}{\sqrt{D_{\text{DA}}}}\right)}{\cD_{i\Omega}\left(\frac{\mu_{\text{DA}}-v_{\text{t}}}{\sqrt{D_{\text{DA}}}}\right)-e^{\Delta}\cD_{i\Omega}\left(\frac{\mu_{\text{DA}}-v_{\text{r}}}{\sqrt{D_{\text{DA}}}}\right)},\label{eq:alpha}\\
\beta(\Omega) & =\frac{ri\Omega(i\Omega-1)}{D(2-i\Omega)}\frac{\cD_{i\Omega-2}\left(\frac{\mu_{\text{DA}}-v_{\text{t}}}{\sqrt{D_{\text{DA}}}}\right)-e^{\Delta}\cD_{i\Omega-2}\left(\frac{\mu_{\text{DA}}-v_{\text{r}}}{\sqrt{D_{\text{DA}}}}\right)}{\cD_{i\Omega}\left(\frac{\mu_{\text{DA}}-v_{\text{t}}}{\sqrt{D_{\text{DA}}}}\right)-e^{\Delta}\cD_{i\Omega}\left(\frac{\mu_{\text{DA}}-v_{\text{r}}}{\sqrt{D_{\text{DA}}}}\right)},\label{eq:beta}
\end{align}
where $\cD_{z}(x)$ is Whittaker's parabolic cylinder function. For
the drift coefficient we need to evaluate Eqs. \eqref{eq:alpha} and
\eqref{eq:beta} at imaginary frequencies. This is convenient, as
it only requires to evaluate $\cD_{z}(x)$ at $z\in\mathbb{R}$. For
the figures, we use the implementation $\mathtt{scipy.special.pbdv(z,x)}$.

\section{Approximation of $D^{(2)}$\label{appendix:Noise-intensity-wdot}}

The diffusion coefficient is the second Kramers-Moyal coefficient\begin{widetext}
\begin{equation}
D^{(2)}(w)=\frac{1}{2}\lim_{\Delta t\rightarrow0}\frac{1}{\Delta t}\left\langle \left[w_{\text{traj}}(t+\Delta t)-w_{\text{traj}}(t)\right]^{2}\right\rangle _{w_{\text{traj}}(t)=w},\label{eq:D2_def-1}
\end{equation}
where $w_{\text{traj}}$ are realizations of \prettyref{eq:model_general}.
We may express the increment as
\begin{equation}
w_{\text{traj}}(t+\Delta t)-w_{\text{traj}}(t)=\int_{t}^{t+\Delta t}dt^{\prime}\dot{w}(t^{\prime})dt^{\prime}.
\end{equation}
Thus,
\begin{equation}
D^{(2)}(w)=\frac{1}{2}\lim_{\tau\rightarrow0}\frac{1}{\Delta t}\int_{t}^{t+\Delta t}dt^{\prime}\int_{t}^{t+\Delta t}dt^{\prime\prime}\left\langle \dot{w}(t^{\prime})\dot{w}(t^{\prime\prime})\right\rangle _{w_{\text{traj}}(t)=w}.\label{eq:km_2_intCvv-1}
\end{equation}
The integration domain is an area of size $\Delta t^{2}$ and we have
a prefactor $1/\Delta t$; thus the only terms in the integrand that
contribute after taking the limit $\lim_{\Delta t\rightarrow0}$ are
those carrying a $\delta(t^{\prime}-t^{\prime\prime})$. Plugging
\prettyref{eq:model_general} into the second moment, we get
\begin{align}
\left\langle \dot{w}(t^{\prime})\dot{w}(t^{\prime\prime})\right\rangle _{w_{\text{traj}}(t)=w} & =\Big\langle\Big[\int_{-\infty}^{0}d\tau\,\kappa(\tau,w)\eta(t^{\prime})x(t^{\prime}+\tau)+\int_{-\infty}^{0}d\tau\,\kappa(-\tau,w)\eta(t^{\prime}+\tau)x(t^{\prime})\Big]\nonumber \\
 & \times\Big[\int_{-\infty}^{0}d\tau\,\kappa(\tau,w)\eta(t^{\prime\prime})x(t^{\prime\prime}+\tau)+\int_{-\infty}^{0}d\tau\,\kappa(-\tau,w)\eta(t^{\prime\prime}+\tau)x(t^{\prime\prime})\Big]\Big\rangle.\label{eq:vel_2nd_moment_expanded-1}
\end{align}
\end{widetext}Multiplying out the product in \prettyref{eq:vel_2nd_moment_expanded-1}
yields four double integrals. Each integral contains a 4-point-correlator
$\left\langle \eta(t_{a})\eta(t_{b})x(t_{c})x(t_{d})\right\rangle $,
which is in general difficult to evaluate. As an approximation, we
assume that we may apply Wick's theorem for the treatment of this
4-point-correlator even though $(\eta,x)$ is not a Gaussian process.
Proceeding with this assumption, we next note that in \prettyref{eq:km_2_intCvv-1},
only terms carrying a $\delta(t^{\prime}-t^{\prime\prime})$ contribute.
Cross-correlations between $x(t_{a})$ and $\eta(t_{b})$ may contain
instantaneous parts $\propto\delta(t_{a}-t_{b})$, however, these
will eliminate \emph{both} integrals in \prettyref{eq:vel_2nd_moment_expanded-1}
and leave no Dirac delta for the integral in \prettyref{eq:km_2_intCvv-1};
such contributions thus vanish at $\Delta t\rightarrow0$. Furthermore,
as argued above, constant parts do not contribute. This leaves us
with 
\begin{equation}
\left\langle \eta(t_{a})\eta(t_{b})x(t_{c})x(t_{d})\right\rangle \approx\nu r(w)\delta(t_{a}-t_{b})\delta(t_{c}-t_{d})
\end{equation}
as the only non-vanishing contribution. The cross products in \prettyref{eq:vel_2nd_moment_expanded-1}
produce constant parts and thus do not contribute. Finally, the diagonal
products yield
\begin{align}
\left\langle \dot{w}(t^{\prime})\dot{w}(t^{\prime\prime})\right\rangle _{w_{\text{traj}}(t)=w} & =\nu r(w)\,\delta(t^{\prime}-t^{\prime\prime})\int_{-\infty}^{\infty}d\tau\,\kappa(\tau,w)^{2}
\end{align}
such that
\begin{align}
D^{(2)}(w) & =\frac{1}{2}r(w)\nu\int_{-\infty}^{\infty}d\tau\,\kappa(\tau,w)^{2}\nonumber \\
 & =\frac{1}{4}r(w)\nu(\Delta_{c}^{2}\tau_{c}+r_{ac}^{2}w^{2}\tau_{ac}).\label{eq:D2_result}
\end{align}

\section{\label{appendix:Breakdown}Breakdown of \prettyref{eq:D1} for low
noise}

\begin{figure*}
\includegraphics{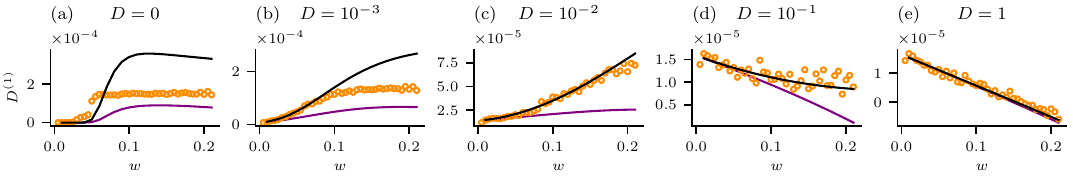}

\caption{\label{fig:Breakdown}Breakdown of \prettyref{eq:D1} at low noise
intensity. Drift coefficient \prettyref{eq:D1} with (black line)
and without (purple line) the impact of the noise response $\beta$,
and simulation results (orange circles). The parameters correspond
to the diffusion coefficient in \prettyref{fig:single_synapse_full}(b),
but the noise intensity $D$ is varied as indicated in the title and
$\mu$ is adjusted such that for $w=0.1$, the firing rate is $r=0.1$.
The integration timestep in the simulation is $dt=10^{-4}$ and the
timestep to evaluate the drift coefficient is $\Delta t=10^{-2}$,
the drift coefficient is averaged over 10k trials.}
\end{figure*}

The LIF neuron \prettyref{eq:lif} studied throughout the manuscript
is driven by white Gaussian noise with intensity $D$, which aims
to model internal and external noise sources. In addition, in the
recurrent network \prettyref{eq:brunel_net_E-1}, the noise intensity
also captures the recurrent input: it is now a sum of non-negative
contributions \prettyref{eq:tot_intput_a} which is typically substantially
large. However, in order to explore the limit case, we now consider
the input current to the neuron to be almost deterministic (small
$D$). In this case, the approximations leading to \prettyref{eq:D1}
and \prettyref{eq:D2} become problematic since the expressions for
the instantaneous firing rate \prettyref{eq:siegert} and for the
response functions \prettyref{eq:alpha} and \prettyref{eq:beta}
rely on the diffusion approximation. In \prettyref{fig:Breakdown},
we demonstrate this breakdown: We study a synapse with weight $w$
driving a neuron \prettyref{eq:lif} with input $\mu+\sqrt{2D}\xi(t)+w\eta(t)$.
We vary the noise intensity $D$ and adjust $\mu$, such that, for
$w=0.1$, the firing rate \prettyref{eq:siegert} of the neuron is
$0.1$ (corresponding to a rate of $5\,\text{Hz}$ for a membrane
time constant of $20\,\text{ms}$). The drift coefficient of the synaptic
weight, \prettyref{eq:D1}, is well captured by the theory for $D>10^{-2}$
but breaks down below. In \prettyref{fig:Breakdown}(b), we observe
that for small enough synaptic weight $w\lesssim0.1$, the theory
works even below $D=10^{-2}$; likely this reflects that for small
amplitudes $w$, the Poissonian input is closer to its diffusion approximation.

\section{Mean-field theory of the recurrent network\label{appendix:MFT_recurrent}}

The drift- and diffusion coefficients of feed-forward synapses in
\prettyref{eq:D1D2_MFT} depend on the input from the recurrent network.
To determine its statistics, we follow \citep{Brunel00_183}; for
more sophisticated approaches to population dynamics see e.g. \citep{Schwalger2017_e1005507}.
The main difference to \citep{Brunel00_183} is that we have three
populations, target, non-target, and inhibitory neurons, and that
for $h\neq1$, inhibitory neurons receive stronger excitation than
excitatory neurons. The total input to excitatory target and non-target
neurons is presented in \prettyref{eq:tot_intput_a}. The total input
to inhibitory neurons is
\begin{align}
\mu_{I}^{\text{tot}} & =\mu_{I}+hJC_{E}r_{E}-gJC_{I}r_{I},\nonumber \\
D_{I}^{\text{tot}} & =D_{I}+\frac{1}{2}(hJ)^{2}C_{E}r_{E}+\frac{1}{2}(gJ)^{2}C_{I}r_{I}.\label{eq:total_input_inh}
\end{align}
Here, the excitatory firing rate is the weighted sum of the target
and the non-target neuron's firing rate
\begin{equation}
r_{E}=f_{s}r_{1}+(1-f_{s})r_{0}.
\end{equation}
Thus, the input to all neurons is determined by the firing rate $r_{1}$,
$r_{0}$, and $r_{I}$, which in turn are determined by \prettyref{eq:siegert}.
This self-consistent set of equations is solved using a damped saddle-point

\end{document}